\documentclass[aps,onecolumn,floatfix,notitlepage]{revtex4-1}
\usepackage{graphicx}
\usepackage[usenames,dvips]{color}

\begin{document}
\title{
Structure and dynamics of ring polymers:
entanglement effects because of solution density and ring topology
}

\author{Angelo Rosa}
\email{anrosa76@gmail.com}
\affiliation{
SISSA - Scuola Internazionale Superiore di Studi Avanzati and IIT - Italian Institute of Technology (SISSA unit) \\
Via Bonomea 265, 34136 Trieste (Italy)
}

\author{Enzo Orlandini}
\email{orlandini@pd.infn.it}
\affiliation{
Dipartimento di Fisica and Sezione INFN,\\ Universit\`a di Padova,
Via Marzolo 8, 35131 Padova (Italy)
}

\author{Luca Tubiana}
\email{tubiana@sissa.it}
\author{Cristian Micheletti}
\email{michelet@sissa.it}
\affiliation{
SISSA - Scuola Internazionale Superiore di Studi Avanzati, Via Bonomea 265, 34136 Trieste (Italy)
}

\date{\today}

\begin{abstract}
The effects of entanglement in solutions and melts of unknotted ring polymers have been addressed by several theoretical and numerical studies.
The system properties have been typically profiled as a function of ring contour length at fixed solution density.
Here, we use a different approach to investigate numerically the equilibrium and kinetic properties
of solutions of model ring
polymers.
Specifically, the ring contour length is maintained fixed,
while the interplay of inter- and intra-chain entanglement is modulated by varying
both solution density (from infinite dilution up to $\approx 40 \%$ volume occupancy)
and ring topology (by considering unknotted and trefoil-knotted chains).
The equilibrium metric properties of rings with either topology are
found to be only weakly affected by the increase of solution density.
Even at the highest density, the average ring size, shape anisotropy
and length of the knotted region differ at most by $40 \%$ from those of isolated rings.
Conversely, kinetics are strongly affected by the degree of inter-chain entanglement:
for both unknots and trefoils the characteristic times of ring size relaxation,
reorientation and diffusion change by one order of magnitude across
the considered range of concentrations.
Yet, significant topology-dependent differences in kinetics are observed
only for very dilute solutions (much below the ring overlap threshold).
For knotted rings, the slowest kinetic process is found to correspond to the diffusion of the knotted region along the ring backbone.
\end{abstract}

\maketitle

\section{Introduction}\label{sec:intro}
Characterizing the equilibrium and kinetic properties of semi-dilute
solutions of cyclic polymers (rings) is one of the major remaining
challenges in theoretical and experimental polymer physics
\cite{rubinstein_ring,catesdeutsch,cates_ring1,cates_ring2,kapnistos2008,kremergrosberg2011_stat,kremergrosberg2011_dyn}.

One aspect of these systems that is very actively investigated regards
how intra-chain and inter-chain entanglement reverberate on
physical properties of rings solutions. Entanglement effects
have been extensively investigated, and are hence well understood, for
linear, i.e. open, polymer chains. However, unlike linear chains,
physical properties of dense solutions of rings strongly depend
on their preparation history during which the ring's geometrical entanglement
is locked permanently in topological constraints such as links and knots.
Because of these constraints, the same concepts
that proved so useful in the characterization of dense solution of linear chains
are usually not easily, nor profitably, transferred to the case of
circular polymers.

A chief example, that paradigmatically highlights the qualitative
differences of systems of closed versus open chains, is offered by the
so-called Edwards-DeGennes {\it reptation} model \cite{doi,degennes71}
for the kinetics of dense linear polymers. According to this model,
each chain is confined inside a tube-like region resulting from the
excluded-volume interactions with the neighboring chains. The motion
of each chain therefore consists of a one-dimensional diffusion along
the tube centerline resulting from the ``inchworm movements'' of small
sub-chain loops. This type of local motion is found also in polymer
rings moving through a fixed array of obstacles \cite{rubinstein_ring}.
However, the asymptotic standard
diffusive behavior of polymer chains in melt depends on the
ongoing process of ``tube renewal'' which results from the fact that
the two polymer ends are practically free and hence capable of
exploring and realizing new configurations
\cite{doi,degennes71,auhl,everaers_science,tzoumanekas2006topological,lahmar2009onset}.
This mechanism, which has
received striking experimental confirmations as being the dominant
kinetic process in dense polymer melts \cite{wisch2003},
explicitly builds on the linear character of the chains in solution.
Hence, it is not applicable, at least in its conventional formulation, to ring polymers.
The latter must consequently relax and
move in dense solution according to different kinetic
mechanisms \cite{catesdeutsch}, which are
only recently being characterized by means of
computationally-intensive molecular dynamics simulations
\cite{Hur_et_al_2006,Tsolou_et_al_2010,Hur_et_al_2011,kremergrosberg2011_stat,kremergrosberg2011_dyn,virnau_mirny,virnau_binder}.

Ring topology affects strongly also equilibrium properties of the chains, and in particular metric ones
\cite{quake,moore2004,kimklein2004,orlandinirev,orlandinistella}.
In fact, the closure condition not only introduces correlations in the
backbone orientation that are of longer range than for linear chains
\cite{witz_2008,druber_et_al_2010}, but it preserves the knotted
topology of the ring.  If chain self-crossing is disallowed
(no bond is cut and rejoined) then the ring knotted or unknotted
state, trapped at the time of circularization, is maintained at all times.

The ring self-entanglement associated to a given topological state is
so important that it affects the scaling of metric properties of {\it isolated} rings,
as first argued by Des Cloizeaux \cite{descloizeaux},
and later by Grosberg and Deguchi \cite{grosberg2009,Matsuda_et_al_2003} and other studies
(see cited references in \cite{micheletti_physrep}).
In particular,
Cates and Deutsch \cite{catesdeutsch} and M{\"u}ller et al. \cite{cates_ring1,cates_ring2}
have suggested that strong inter-chain topological interactions lead to
pronounced chain compaction of {\it unlinked} and {\it unknotted}
rings in semi-dilute solutions, and that the typical chain
configuration should resemble a branched polymer (akin to a lattice
animal) \cite{cates_ring1,cates_ring2}.

These seminal studies spurred several theoretical and computational
efforts where equilibrium and kinetic properties of ring solutions
were studied at fixed density -- almost invariably corresponding to
melt conditions -- and for increasing ring length. These approaches,
have provided considerable insight into the expected asymptotic
scaling of the ring's metric properties
\cite{kremergrosberg2011_stat,kremergrosberg2011_dyn,virnau_mirny,virnau_binder}.
In particular, recent evidence
\cite{kremergrosberg2011_stat,kremergrosberg2011_dyn} suggests that
the branched polymer regime is likely at a crossover between the
Gaussian regime for short rings and the ``crumpled globule'' regime
for long ones \cite{grosberg}.  In the latter regime, each ring
portion is highly compact and minimally intermingling with the rest
of the chain \cite{grosberg}, similarly to dense systems of long
biopolymers in vivo, such as eukaryotic chromosomes \cite{plospaper}.

Compared to these studies, here we use
stochastic molecular dynamics simulations
to investigate complementary aspects of the effects that intra- and
inter-chain topological constraints have on kinetic and equilibrium
properties of solutions of model ring polymers.
Specifically, we consider systems
in which the contour length of the rings is fixed while the solution
density is varied so to cover an appreciable range of inter-chain
entanglement. The study is carried out for monodisperse solutions of
unknotted rings as well as trefoil-knotted ones. To have a better
insight into topological effects on the statics and dynamics of
knotted rings we look also at the knotted portion of the ring. This is
in general a formidable task and we succeeded in tackling it by
relying on a recently-introduced effective and transparent procedure
for locating knots \cite{tubiana1,tubiana2}.
To the best of our knowledge this is the first time that a systematic off-lattice study
of kinetic and equilibrium properties of rings with non-trivial
topology and of their knotted portions has been carried out.

By first exploring the metric and shape properties of equilibrated
rings in solution we find that their behavior depends appreciably on
ring topology and that not only it differs from the one seen for
linear polymers in solutions but even from other physical realizations
of dense ring polymers.
Specifically, unlike what has been observed in
confined \cite{micheletti_pnas,tubiana1,tubiana2} and collapsed knotted
rings \cite{marcone2007,baiesi2011}, the average size of the knotted
portion of a ring in solution is only weakly affected by the increase (for
increasing monomer concentration) of the system's geometrical entanglement.
In particular, by increasing the solution density,
no crossover to delocalized knots and no multiscale behavior of the
entanglement is observed, unlike the case of spherical confinement \cite{tubiana2}.

The presence of topological constraints makes the dynamics of rings
in solution an even richer phenomenon whose understanding requires a
full knowledge of the relationship between the spatial motion of a
ring in a melt and the motion of its knotted portion along the ring backbone.
By using the knot location tool previously mentioned, and
considering dynamic observables commonly used in linear polymer
contexts we show that several time scales are at play:
the autocorrelation time of ring configurations, the time required by the
rings to diffuse over regions comparable to their average size and -- for knotted rings --
the time required by the knot to diffuse over the ring contour.

Various relaxation properties of the entire ring are found to change
by one order of magnitude across the considered density range.
The slowest kinetic process is associated with the diffusion along the
backbone of the knotted region.
This property, that to the best of our knowledge has not been pointed out before,
seamlessly integrates with the other kinetic aspects thus offering a consistent picture for polymer
relaxation in solutions of topologically constrained rings.

\section{Model and methods}\label{sec:modmeth}

\subsection{The model}\label{sec:model}
To model the rings in solution, we use the bead-spring polymer model introduced by Kremer and Grest \cite{kremer_jcp}.
The model accounts for the connectivity, bending rigidity, excluded volume
and topology conservation of polymer chains.

Specifically, the intra-chain energy consists of the following terms:
\begin{eqnarray}
{\cal H}_{intra} & = & \sum_{i=1}^n [ U_{FENE}(i, i+1) + \\
                 &   & U_{br}(i, i+1, i+2) +\\
                 &   & \sum_{j=i+1}^n U_{LJ}(i,j) ]
\end{eqnarray}
where $n$ is the total number of beads per ring, and $i$ and
$j$ run over the indices of the beads. The latter are assumed to be
numbered consecutively along the ring from one chosen reference monomer.
The modulo-$n$ indexing is implicitly assumed because of the ring periodicity.  

From now on we shall take the nominal bead diameter, $\sigma$, as
the unit length and adopt the following notation: the position of the
center of the $i$th beads is indicated by $\vec{r}_i$ while the
pairwise vector distance of beads $i$ and $j$ is denoted as
$\vec{d}_{i, j} = \vec{r}_j - \vec{r}_i$ and its norm simply as $d_{i, j}$.

With this notation the chain connectivity term,
$U_{FENE}(i,i+1)$ is expressed as:
\begin{equation}\label{eq:fenepot}
U_{FENE}(i,i+1) = \left\{
\begin{array}{l}
- {k \over 2} \, R^2_0 \, \ln \left[ 1 - \left( {d_{i,i+1} \over R_0} \right)^2 \right], \, d_{i,i+1} \leq R_0\\
0, \, d_{i,i+1} > R_0
\end{array}
\right.
\end{equation}
where $R_0=1.5 \sigma$, $k=30.0 \epsilon / \sigma^2$ and
the thermal energy $k_B\, T$ equals $1.0 \epsilon$ \cite{kremer_jcp}.
The bending energy has instead the standard Kratky-Porod form (discretized worm-like chain):
\begin{equation}\label{eq:stiffpot}
U_{br}(i, i+1 ,i+2) = \frac{K_B\, T \, \xi_p}{\sigma}  \left (1 - \frac{{\vec d}_{i,i+1} \cdot  {\vec d}_{i+1,i+2}}{d_{i,i+1} \, d_{i+1,i+2}} \right )
\end{equation}
where $\xi_p=4.5 \sigma$ is the nominal persistence length of the chain \cite{rubinstein_colby}.
Polymer chains are significantly bent by thermal fluctuations at contour lengths larger than $\ell_K$,
where $\ell_K = 2 \xi_p = 9.0\sigma$ is the Kuhn length of the chain \cite{doi,rubinstein_colby}.

The excluded volume interaction between distinct beads (including consecutive ones)
corresponds to a purely repulsive Lennard-Jones potential:
\begin{equation}\label{eq:ljpot}
U_{LJ}(i,j) = \left\{
\begin{array}{l}
4 \epsilon [(\sigma/d_{i,j})^{12} - (\sigma/d_{i,j})^6 + 1/4], d_{i,j} \leq \sigma 2^{1/6}\\
0, d_{i, j} > \sigma 2^{1/6}
\end{array}
\right. .
\end{equation}
This repulsive interaction controls the inter-chain excluded volume too:
\begin{equation}
{\cal H}_{inter} = \sum_{I=1}^{N-1} \sum_{J=I+1}^N U_{LJ}(i,j)
\end{equation}
where $N$ is the number of rings in solution and the index
$i$ $[j]$ runs over the beads in chain $I$ $[J]$.

\subsection{Simulation details}\label{sec:equil}
We consider solutions of $N = 64$ rings, each consisting of $n = 216$ beads
(i.e., the ring contour length, $L_c = 216 \sigma$ corresponds to $24 \ell_K$)
at six different monomer densities,
$\rho$: $\rho \sigma^3 = 0.010, 0.025, 0.050 < \rho^{*} \sigma^3$,
$\rho \sigma^3 = 0.100$ and $\rho \sigma^3 = 0.200, 0.400 > \rho^{*} \sigma^3$.
Notice, that the chosen densities cover the
cross-over to the overlap monomer concentration, $\rho^{*} \sigma^3
\approx 0.1$ (see Eq. \ref{eq:overdens} below).
For reference, we also consider isolated rings,
and solutions of $N=21$ rings in very dilute conditions ($\rho \sigma^3 = 0.003$).

The equilibrium and kinetic properties of these systems are studied
using fixed-volume and constant-temperature Molecular Dynamics (MD) simulations.
The system dynamics is integrated with the LAMMPS engine \cite{lammps}
with Langevin thermostat (target temperature $= 1.0$ LJ-units).
The elementary integration time step is chosen equal to $\Delta t = 0.012\tau_{MD}$,
where $\tau_{MD}=\sigma(m/\epsilon)^{1/2}$
is the Lennard-Jones time, $m$ is the bead mass and the friction
coefficient, $\gamma$, corresponds to $\gamma/m = 0.5 \tau_{MD}^{-1}$ \cite{kremer_jcp}.
We stress that,
because we are dealing with model polymers, the solvent is not explicitly included in our simulation.
The polymer-solvent interaction is effectively accounted for
only through the Langevin thermostat
and related friction term.

{\it Preparation of initial configurations} -- At each density, we
consider monodisperse solutions, consisting either of unknotted {\it
  or} trefoil-knotted rings.  In the following we shall often use the
standard shorthand notation UN and $3_1$ to denote unknotted and
trefoil-knotted topologies, respectively.

\begin{figure*}
\includegraphics[width=6.0in]{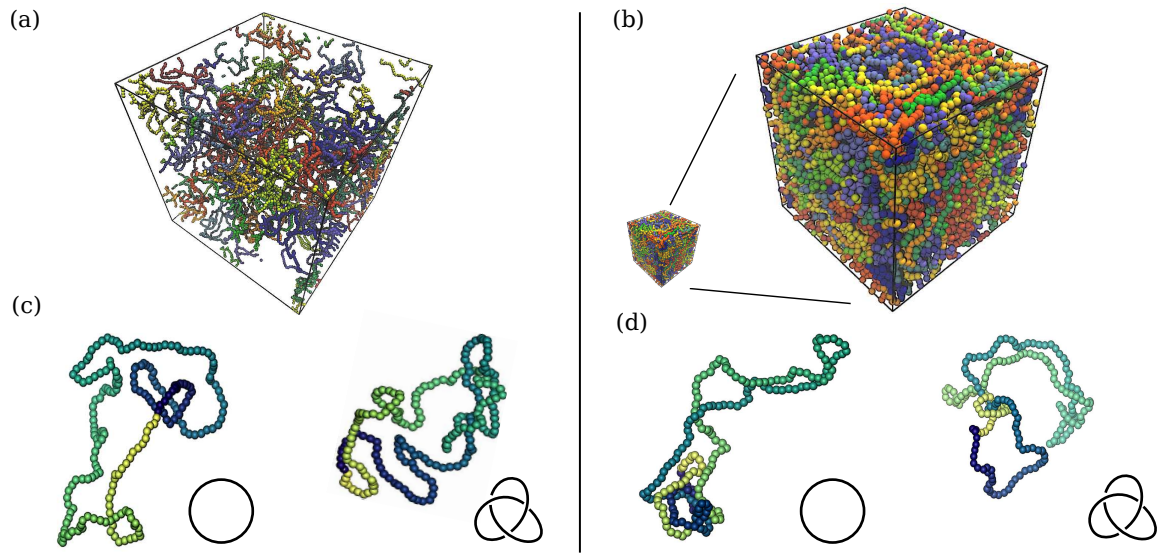}
\caption{Typical configurations of ring polymer solutions, at $\rho\sigma^{3}=
  0.025$ (a) and $\rho\sigma^{3}  = 0.4$ (b), and
  corresponding selected rings with trefoil-knotted (c and d, right)
  and unknotted topologies (c and d, left).}
\label{fig:melt}
\end{figure*}

The system initialization consists of placing a template
unknotted or trefoil-knotted ring at the center of a cubic cell with
periodic boundary conditions. The template configuration is replicated
4 times along each spatial direction.
The linear dimension of the cubic cell is large enough to avoid overlap and linking between the
template copies and yields an initial monomer density of about
$\rho \sigma^3 \approx 5 \cdot 10^{-3}$.
Prior to the production runs, the cell is first evolved at constant pressure until the desired target
density is reached and is subsequently equilibrated for a time span of about $10^5 \tau_{MD}$.
At all considered densities, this time span exceeds the time required by one ring to diffuse over distances larger
than its typical size.
Production runs have a typical duration of $6 \cdot 10^6 \tau_{MD}$.
Typical equilibrated configurations at two different densities are shown in Fig. \ref{fig:melt}.

\subsection{Ring shape and size}
As customary, the salient metric properties (shape and size) of a
given ring are characterized through the gyration tensor, ${\mathbf Q}$.
The entries of this $3 \times 3$ matrix are given by:
\begin{equation}\label{eq:gyrtensdef}
Q_{\alpha, \beta} =
\frac{1}{n} \sum_{i=1}^{n}
({\vec r}_{i, \alpha} - {\vec r}_{CM, \alpha}) ({\vec r}_{i, \beta} - {\vec r}_{CM, \beta}) ,
\end{equation}
where $\alpha$ and $\beta$ run over the three Cartesian components and
${\vec r}_{CM} = \frac{1}{n} \sum_{i=1}^{n} {\vec r}_i$ is the spatial
location of the ring center of mass.
The non-negative eigenvalues of $\mathbf Q$, ranked with decreasing magnitude, $\Lambda_1$, $\Lambda_2$ and
$\Lambda_3$ correspond to the square length of the principal axes of
the ring gyration ellipsoid.  Accordingly, their relative magnitude
conveniently captures the overall spatial anisotropy of the rings,
while their sum yields the ring square radius of gyration:
\begin{equation}\label{eq:gyrrad}
R_g^2 =  Tr {\mathbf Q}  = \sum_{\alpha = 1}^3  \Lambda_{\alpha}  ,
\end{equation}
where $Tr$ is the {\it trace} operator.
To measure the typical size and anisotropy of the ensemble of rings in
solution, we compute the averages $\langle R_g^2 \rangle$ and $\langle
\Lambda_{i=1, 2, 3} \rangle$ where the brackets $\langle ... \rangle$
denote averaging over all rings of all system
snapshots at a given solution density.

\subsection{Identifying the knotted portion of the rings}\label{sec:knotid}

Because of the chain connectivity constraint, the global topological
state of the rings is preserved in the course of the MD evolution. One
way to characterize the interplay of geometrical and topological
entanglement is by establishing the magnitude of the typical length of
the ring portion that accommodates the knot \cite{katritch2000,marcone2007,tubiana1}.

\begin{figure}
\includegraphics[width=3.0in]{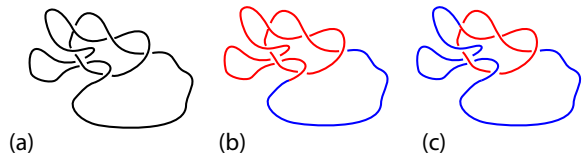}
\caption{
Knotted portions in a non-trivially embedded trefoil knot (a), as identified by two different search approaches.
(b)
Top-down approach: the red arc is the shortest continuously knotted portion.
It can not be shortened without loosing the $3_1$ topology.
(c)
Bottom-up approach: the red is the shortest arc with $3_1$ topology.
Complementary arcs (in blue) in both approaches must be unknotted (see Ref. \cite{tubiana1}, for details).
}
\label{fig:trovailnodo}
\end{figure}

From a mathematical point of view, the problem of locating the knot in
a closed chain is ambiguous \cite{millett_stasiak,virnau_kantor,micheletti_physrep}.
The difficulties are essentially two \cite{tubiana1}.
First, to assess if the knot is contained in a given ring portion,
which is an open arc, it is necessary to define a procedure for
closing it by suitably bridging the arc ends. In fact, only after the
arc is closed it is possible to characterize its topological state by
using topological invariants.  Viable closure procedure should
introduce the least possible additional entanglement.  This criterion
is well satisfied by the minimally-interfering closure that some of us
introduced recently \cite{tubiana1} and that will accordingly be used
throughout our study.
The second difficulty lies in the fact that the geometrical
entanglement of a knotted ring can be so high that the identification
of the region that accommodates the knot depends on the method used to search for it \cite{tubiana2}.
Arguably, upper and lower bounds to the
length of the region accommodating the knot are offered by the
top-down and bottom-up searches illustrated in
Fig. \ref{fig:trovailnodo}.
Both methods were applied here and systematically yielded consistent
results, see Supporting Fig. 1. 

Based on the robustness of the knot localization procedure for rings in solution,
we shall limit considerations to the bottom-up search.
This method returns the shortest arc having the same topology of the whole ring (a
trefoil knot, in our case) while its complementary arc is unknotted,
see Fig. \ref{fig:trovailnodo}c.

\subsection{Surface accessible area}\label{sec:sas}
The rings surface accessible area was computed using the SAS
routine of the GROMACS package \cite{gromacsas,gromacs4} using a probe
sphere of diameter equal to the bead diameter, $\sigma$.  The computed
surface area includes contributions from inner cavities of the ring
conformations that are large enough to accommodate the probe sphere.

\section{Results and discussion}\label{sec:results}

\subsection{Geometric properties of the rings}\label{sec:geomprops}
We first report on how various equilibrium metric properties of the
rings depend on their topology and solution density.
In a progression from global to finer aspects, we shall consider the rings
size, shape, exposed surface area, the mean-square distance of points
at increasing arclength separation and the degree of localization of
the knot that they accommodate (in case of non-trivial topology).
 
The rings consist of $n=216$ beads of diameter $\sigma$ with
associated Kuhn length, $\ell_K = 9\sigma$. The considered monomer
density of the solution, $\rho\sigma^{3}$, spans the interval $[0.01,
0.4]$.  This range spans from a dilute situation to one where the
inter-chain entanglement is significant and yet the solution is
isotropic \cite{uchida}.
The lowest monomer density is much smaller than the one at
which one expects significant ring overlap.
The overlap density is estimated as \cite{doi},
\begin{equation}
\rho^* \approx \frac{L_c / \sigma}{\langle R_g^2 \rangle ^{3/2}} \ .
\label{eq:overdens}
\end{equation}
where $\langle R_g^2 \rangle$ is the mean square radius of gyration of an isolated ring.
Using the Zimm-Stockmayer estimate for the isolated ring size \cite{zimm_stock},
$\langle R_g^2 \rangle = { L_c\, \ell_K \over 12} \approx 162 \sigma$ yields $\rho^{*}\sigma^3 \approx 0.1$, which is
an order of magnitude larger than the minimal monomer density considered here.

Above this density, significant mutual entanglement of the chains is expected.
For dense solutions of {\it linear} polymers, the intricacy
of the melt \cite{degennes71,doi} and the resulting physical
properties \cite{everaers_science,uchida} are captured by the entanglement length, $L_e$.
Roughly speaking, this quantity corresponds to the typical chain arclength separation
between two consecutive topological constraints (known as {\it entanglements})
arising from inter-chain {\it un}crossability.
It has been recently proposed \cite{uchida} that the relationship tying $L_e$, $\ell_K$ and the solution
density, $\rho$ is adequately captured by the following
phenomenological expression:
\begin{equation}\label{eq:ralf}
\frac{L_e}{\ell_K} \approx (0.06 \, (\rho \, \sigma \, \ell_K^2))^{-2/5} + (0.06 \, (\rho \, \sigma \, \ell_K^2))^{-2} .
\end{equation}
Using the previous expression, one has that the interval of
$\rho$ considered here corresponds to a wide range of $L_e$ for the
equivalent system of linear chains. Indeed, for $\rho\sigma^3$ going
from $10^{-2}$ to $0.4$ we have that $L_e / \ell_K$ spans from
$\approx 400$ to $\approx 1$.
In the latter situation, the entanglement length is approximately equal to the Kuhn length,
and hence the system is at the crossover from the loosely- to the tightly-entangled
regimes (i.e. the chain is nearly straight between two consecutive entanglements \cite{morse}).
Higher densities are not considered, because for $\rho \sigma^3 \gtrsim 0.5$ the onset of the Onsager
isotropic/nematic transition \cite{uchida} is expected to break the
spatial isotropy of the solution.

\begin{table}[htp]
\begin{center}
\begin{tabular}{|c|c||c|c|c|}
\hline
{}  &  $ \langle R_g^2 \rangle$ $ [\sigma^2] $  &  $ \langle \Lambda_1 \rangle$ $[\sigma^2] $  &  $ \langle \Lambda_2 \rangle $ $ [\sigma^2] $  &  $ \langle \Lambda_3 \rangle $ $ [\sigma^2] $ \\
\hline
UN     &  $ 175.4 \pm 0.6 $  &  $ 116.3 \pm 0.6 $  &  $ 43.9 \pm 0.1 $  &  $ 15.2 \pm 0.1 $ \\
$3_1$  &  $ 120.8 \pm 0.7 $  &  $  77.9 \pm 0.7 $  &  $ 30.6 \pm 0.1 $  &  $ 12.3 \pm 0.1 $ \\
\hline
FJR    &  $ 168.7 \pm 0.4 $  &  $ 108.5 \pm 0.4 $  &  $ 42.1 \pm 0.2 $  &  $ 18.1 \pm 0.1 $ \\
\hline
GR     &  $ 161.6 \pm 0.7 $  &  $ 106.1 \pm 0.6 $  &  $ 39.7 \pm 0.1 $  &  $ 16.2 \pm 0.1 $ \\
\hline
\end{tabular}
\caption{
\label{tab:isolrings}
Average value of the mean square radius of gyration and eigenvalues of
the gyration tensor for unknotted and trefoil-knotted isolated rings.
Analogous quantities are also shown for
an equivalent freely-jointed ring (FJR) and Gaussian ring (GR) of 24 bonds.
The FJR bond length and the GR root-mean square bond length are equal to $\ell_K = 9 \sigma$.
}
\end{center}
\vspace{-0.6cm}
\end{table}

Following the outline set at the beginning of this section we first
discuss how overall ring metric properties depend on the monomer
density, $\rho$.

For reference, we provide in Table \ref{tab:isolrings} the
average values of the square radius of gyration, $\langle R^2_g
\rangle$, and the eigenvalues of the gyration tensor for isolated
unknotted and trefoil-knotted rings.
Analogous quantities are reported for an equivalent freely-jointed ring (FJR)
and Gaussian ring (GR) of $L_c / \ell_K = 24$ bonds \cite{doi}.
The FJR data are obtained by a Markovian exploration of FJRs,
while the GR data are obtained by stochastic molecular dynamics simulations.
Table \ref{tab:isolrings} conveys the effect that intra-chain constraints
(excluded volume and fixed topology) have on the overall size and
shape of rings at infinite dilution.
Both types of constraints are absent in the FJR and the GR.
Notice, that the FJR and GR quantities are typically within $10\%$ of those of unknotted rings.
The difference with analogous quantities for trefoil-knotted rings is substantially larger.
These results reflect the fact that, for the
considered values of $L_c$ and $\ell_K$, unknotted rings dominate the
equilibrium ensemble of infinitely-thin (FJR) or Gaussian
rings with unrestricted topology.
Trefoil-knotted rings therefore possess a degree of
entanglement that would be atypical in rings circularized in
equilibrium.  The presence of the non-trivial topological constraint
causes such rings to be tighter and
slightly more isotropic than unknotted ones. This result is
consistent with the intuitive notion that a certain arclength of
trefoil-knotted rings is ``used up'' \cite{millett_stasiak} in the
non-trivial entanglement (in fact the minimal ropelength required to
tie trefoil-knotted rings is larger than for unknotted ones).

\begin{table*}[htp]
\begin{center}
\begin{tabular}{|c||c|c|c||c|c|c||}
\hline
{} &  UN  &  UN  &  UN  &  $3_1$  &  $3_1$  &  $3_1$\\
Monomer density, $\rho\sigma^{3}$  &  $ \langle R_g^2 \rangle \, [\sigma^2]$  &  $ \langle \Lambda_{1} \rangle / \langle \Lambda_{3} \rangle $ & $ \langle \Lambda_{2} \rangle / \langle \Lambda_{3} \rangle $  &  $ \langle R_g^2 \rangle \, [\sigma^2]$  &  $ \langle \Lambda_{1} \rangle / \langle \Lambda_{3} \rangle $ & $ \langle \Lambda_{2} \rangle / \langle \Lambda_{3} \rangle $\\
\hline
0.010 &  $ 165.59 \pm 0.37 $  &  $ 7.45 \pm 0.05 $  &  $ 2.77  \pm 0.01  $  &  $ 116.07 \pm 0.01 $  &   $ 6.118 \pm 0.001 $  &  $ 2.4151 \pm 0.0004 $\\
\hline                                                                                                              
0.025 &  $ 150.55 \pm 0.39 $  &  $ 7.21 \pm 0.03 $  &  $ 2.63  \pm 0.02  $  &  $ 108.23 \pm 0.08 $  &   $ 5.91  \pm 0.01  $  &  $ 2.335  \pm 0.003  $\\
\hline                                                                                                              
0.050 &  $ 134.64 \pm 0.06 $  &  $ 7.07 \pm 0.02 $  &  $ 2.530 \pm 0.003 $  &  $  97.26 \pm 0.05 $  &   $ 5.80  \pm 0.01  $  &  $ 2.266  \pm 0.006  $\\
\hline                                                                                                              
0.100 &  $ 115.63 \pm 0.09 $  &  $ 7.00 \pm 0.03 $  &  $ 2.44  \pm 0.07  $  &  $  85.62 \pm 0.19 $  &   $ 5.95  \pm 0.02  $  &  $ 2.244  \pm 0.004  $\\
\hline                                                                                                              
0.200 &  $  98.64 \pm 0.10 $  &  $ 7.02 \pm 0.02 $  &  $ 2.385 \pm 0.006 $  &  $  77.94 \pm 0.22 $  &   $ 6.29  \pm 0.03  $  &  $ 2.260  \pm 0.008  $\\
\hline                                                                                                             
0.400 &  $  86.32 \pm 0.85 $  &  $ 7.46 \pm 0.13 $  &  $ 2.41  \pm 0.01  $  &  $  70.11 \pm 0.06 $  &   $ 6.84  \pm 0.03  $  &  $ 2.318  \pm 0.008  $\\
\hline
\end{tabular}
\caption{
\label{tab:gyrrad}
Mean square radius of gyration, $\langle R_g^2 \rangle$ and average
shape of the simulated rings expressed as the ratios $\langle
\Lambda_{1} \rangle / \langle \Lambda_{3} \rangle$ and $\langle
\Lambda_{2} \rangle / \langle \Lambda_{3} \rangle$, where $\langle
\Lambda_{\alpha=1,2,3} \rangle$ are the three average eigenvalues of
the gyration tensor, Eq. \ref{eq:gyrtensdef}.
For very dilute conditions ($\rho \sigma^3 = 0.010$), the calculated quantities are close to
the theoretical values for ideal semi-flexible rings (see Table \ref{tab:isolrings}).
}
\end{center}
\vspace{-0.6cm}
\end{table*}

{\it Ring size.}
Table \ref{tab:gyrrad} reports the same metric properties of Table
\ref{tab:isolrings}, but calculated for rings in solution.
The data suggest that, for both topologies, the ring size decreases with the monomer density,
$\rho$. Specifically, going from $\rho \sigma^3 = 10^{-2}$ to $\rho \sigma^3 = 0.4$ it is seen
that $R_g^2$ decreases from the typical value of isolated rings
down to about half of it. 
The decreasing trend of
the average extension results from the non-concatenation constraint of the rings,
consistent with previous numerical findings \cite{cates_ring1,cates_ring2}.
It should also be noted that the monomer density
attained by each ring is much smaller than the density of the entire solution.
In fact,
even in the densest case $\rho \sigma^3 = 0.4$
the individual ring density --
defined as the number of ring monomers divided by the volume of the average gyration ellipsoid --
is only $\approx 0.05 \sigma^{-3}$ for unknots and $\approx 0.06 \sigma^{-3}$ for trefoil-knotted rings.
These findings suggest that an increase of solution density
promotes extensive chain intermingling as opposed to tight
compaction of the individual rings.

{\it Ring shape.}
A further indication that an increase of the
solution density affects only weakly the metric properties of each ring is
given by looking at the typical ring shape.
For this purpose, following previous studies of either closed or open chains
\cite{bishop,rudnick,alimfrey}, we calculate the average eigenvalues $\langle
\Lambda_{\alpha=1, 2, 3} \rangle$ of the gyration tensor,
Eq. \ref{eq:gyrtensdef} which are shown in Table \ref{tab:gyrrad}.
We note that the ratios $\langle \Lambda_1
\rangle / \langle \Lambda_3 \rangle$ and $\langle \Lambda_2 \rangle /
\langle \Lambda_3 \rangle$ fluctuate by a few percent throughout the
explored range of solution densities and remain close to the overall
anisotropy expected for equivalent, isolated FJRs (see Table \ref{tab:isolrings}).

{\it Exposed surface.}
To complete the statistical characterization of ring conformations 
we next focus on the average exposed surface area of each ring.
This is an important geometric indicator that aptly complements
the radius of gyration \cite{cates_ring1,cates_ring2,kremergrosberg2011_stat,kremergrosberg2011_dyn,grosberg}
in describing the overall degree of ring compactness attained in denser and denser solutions.
We stress that, because the purpose is to characterize the properties
of individual rings, the exposed surface per bead is calculated
separately for each ring, i.e. without taking into account the burial
effect due to the surrounding chains.
Furthermore, the exposed surface is calculated by taking into account voids (cavities) possibly
present in a ring (see Methods, Sec. \ref{sec:sas}).

\begin{figure*}
$$
\begin{array}{cc}
a)
\includegraphics[width=2.4in]{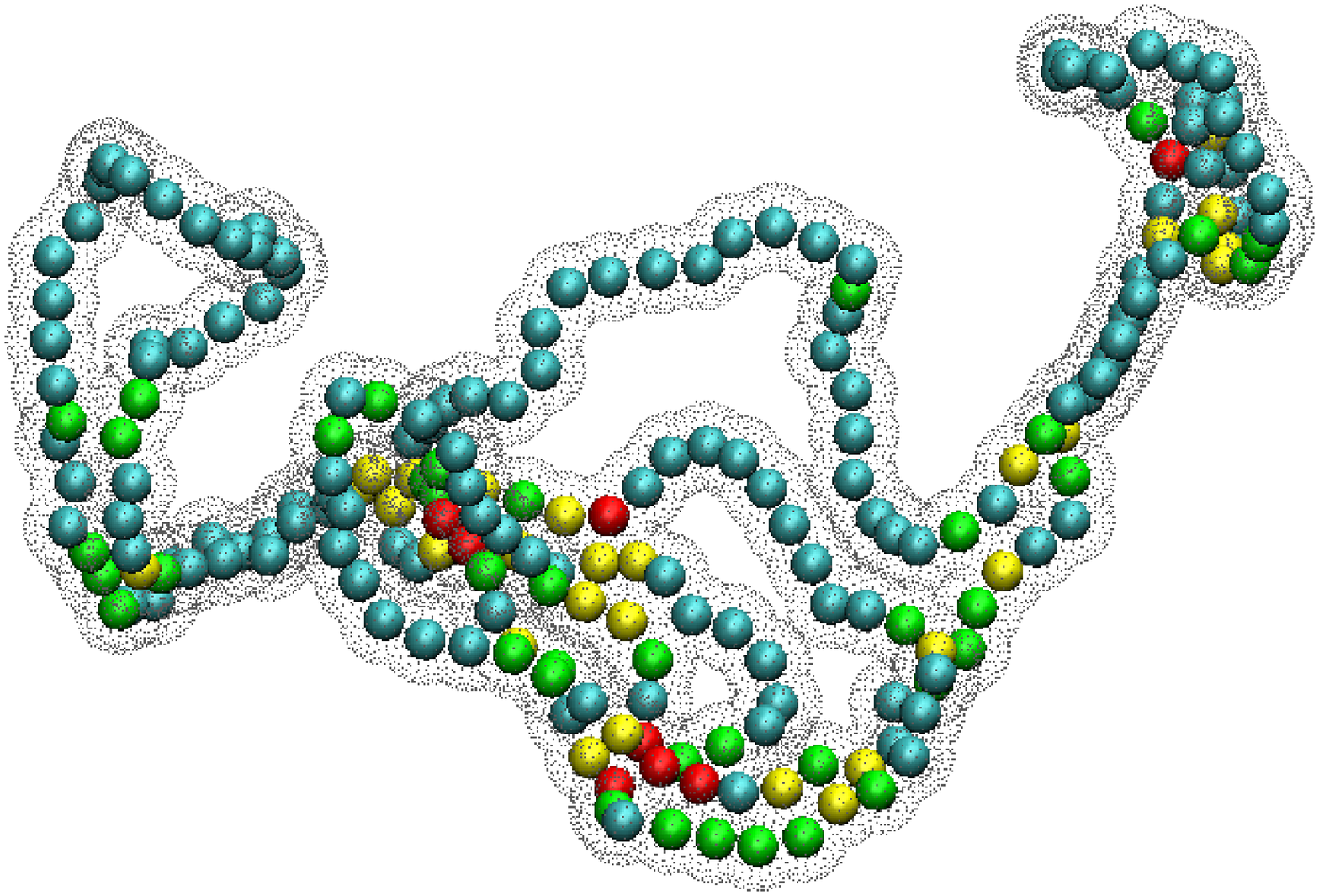} &
b)
\includegraphics[width=3.0in]{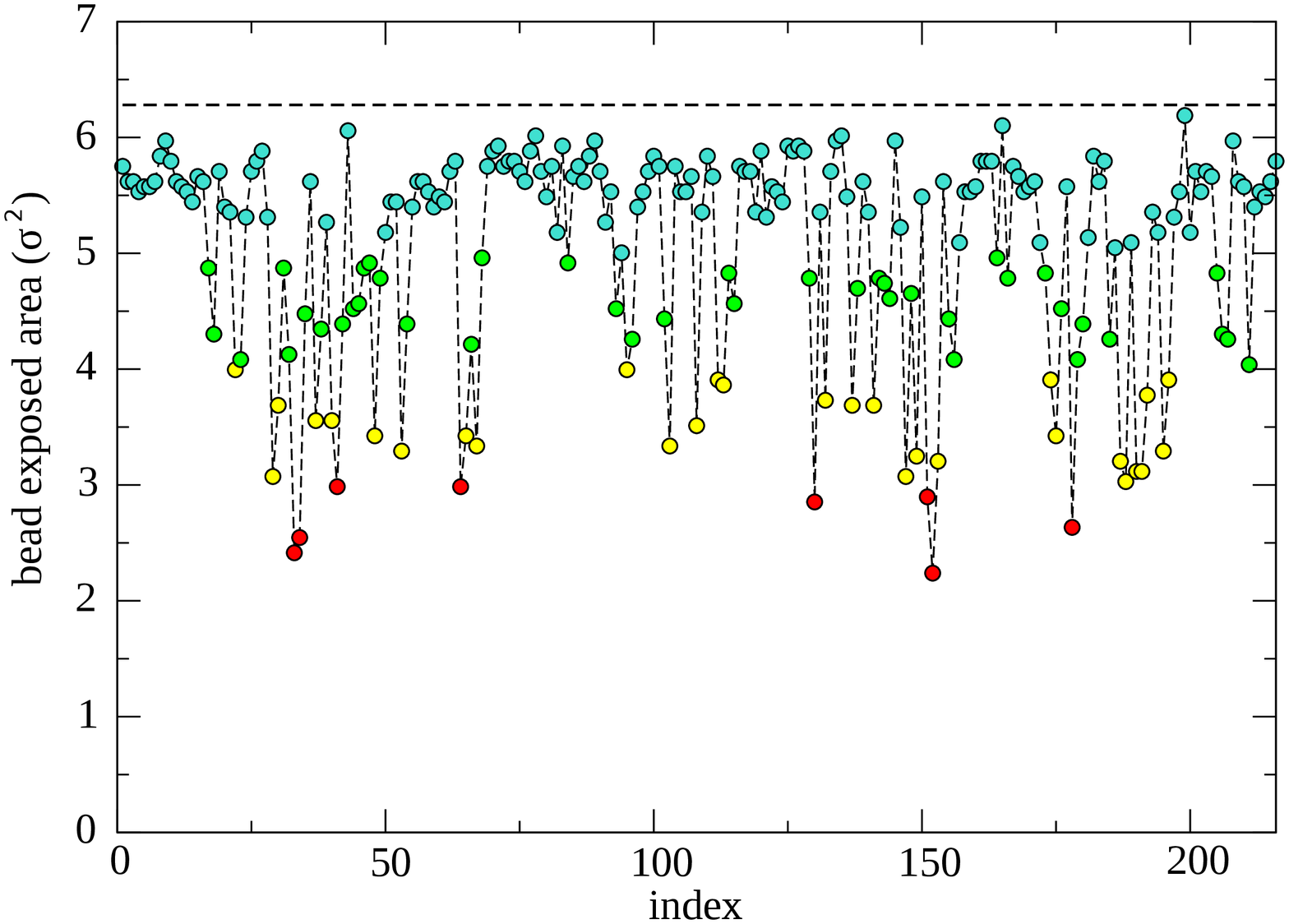} \\
c)
\includegraphics[width=3.0in]{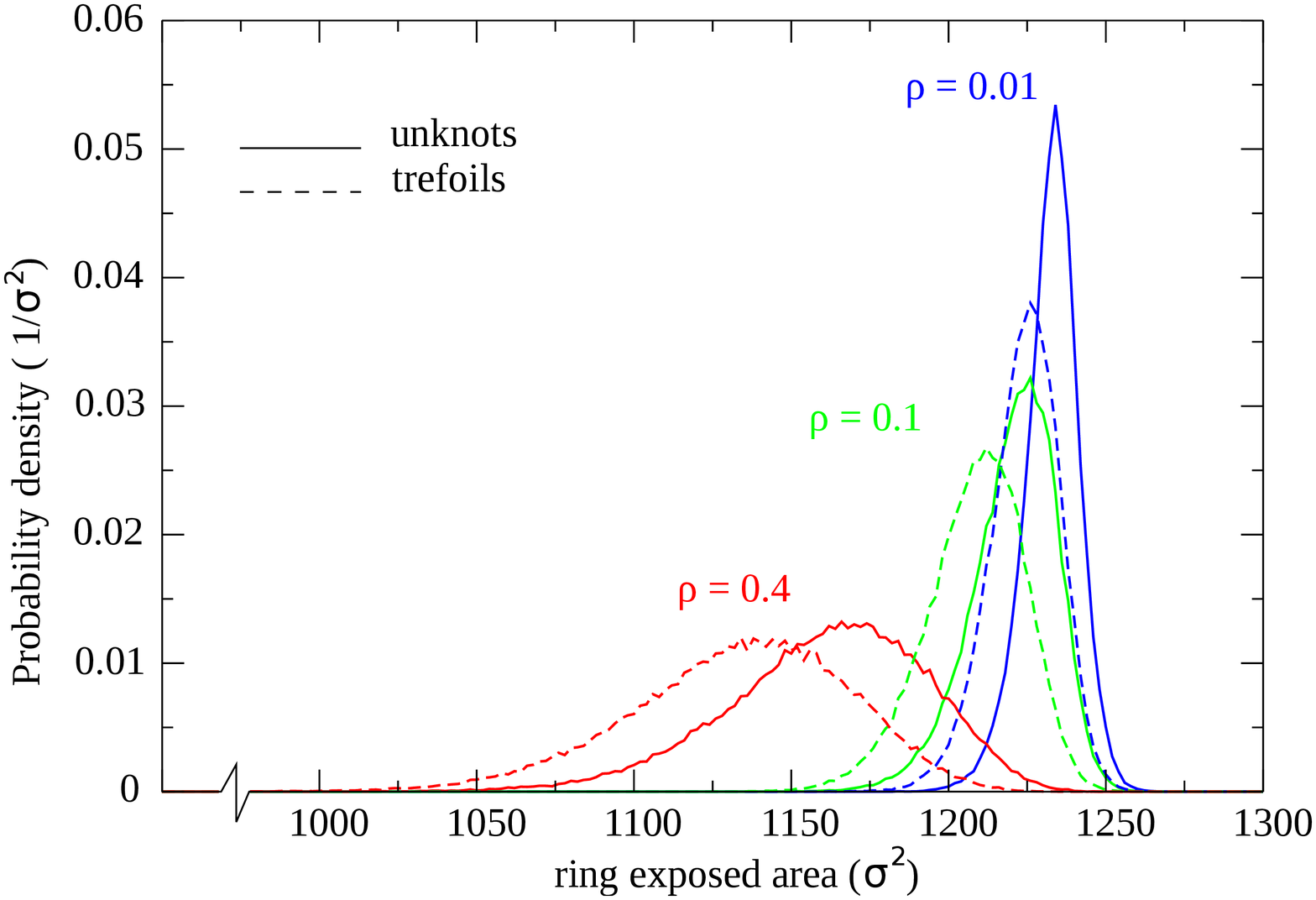} &
d)
\includegraphics[width=3.2in]{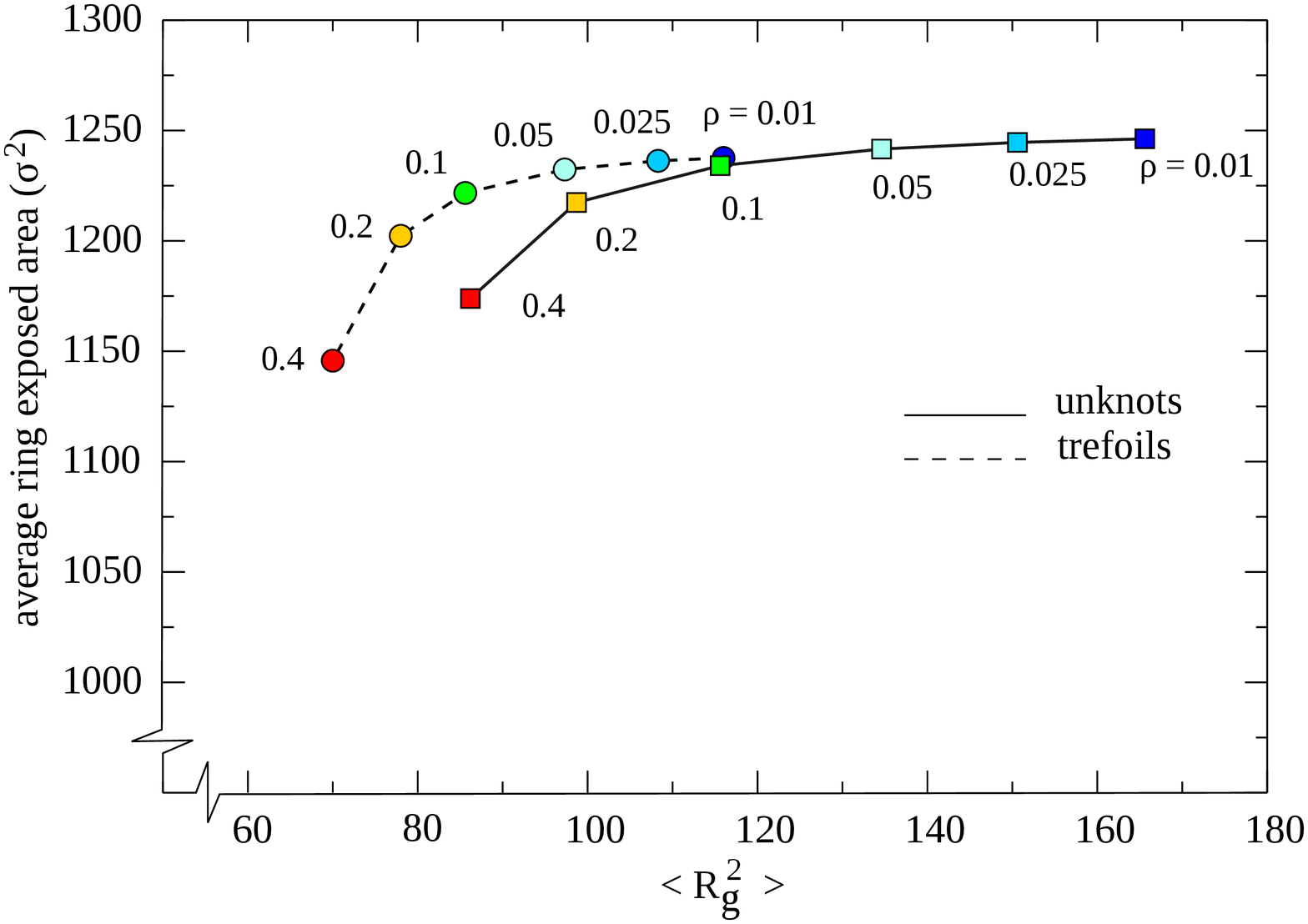}
\end{array}
$$
\caption{
(a)
The shaded region outlines the accessible surface of the shown ring configuration.
This is calculated using the SAS routine of the GROMACS package \cite{gromacs4,gromacsas}
with a probe sphere of diameter equal to the bead diameter, $\sigma$.
Monomer beads are colored from red to cyan for increasing degree of their surface exposure,
shown in panel (b) with the same coloring scheme.
(c)
Probability distribution functions of the total accessible surface per ring,
for different solution densities and chain topologies.
(d)
Average accessible surface per ring as a function of the ring average size,
for different solution densities and chain topologies.
\label{fig:1000asa}
}
\end{figure*}

The varying degree of exposure of the beads in a ring is illustrated
in Fig. \ref{fig:1000asa}b: the profile pertains to the unknotted ring
conformation in panel (a) (picked at $\rho \sigma^3=0.4$), where the
beads are color coded according to the degree of exposure.
For reference, in panel (b) it is also shown the value of the exposed
surface per bead in a perfectly straight chain configuration. For the
case shown in panels (a-b), highly exposed beads nearly approach this
reference value, while the most buried ones have an exposed surface
equal to about $1/3$ of the reference value.
The distributions of the total exposed surface per ring are shown in panel (c):
in particular, we notice that the ring exposed surface covers a moderately narrow interval.
Nevertheless, the exposed area for trefoil-knotted rings is generally smaller than for the
unknots (at the same monomer concentration, $\rho$),
and the mean value of the exposed surface per bead
(averaged over all beads of all rings in various configurations at fixed density)
is a decreasing function of solution density.

Finally, it is interesting to correlate the decrease of the surface area
with the decrease of the ring radius of gyration observed for increasing density.
This relationship is shown in panel (d) of Fig. \ref{fig:1000asa}.
By taking into account the different
scales and offsets of the two axes in the graph
it is realized that unknotted and trefoil rings at the same monomer concentration differ
appreciably by average size (see also Table \ref{tab:gyrrad})
though not by exposed surface.
Secondly, the density-dependent decrease of the radius of gyration is not
paralleled by an analogous decrease of the exposed surface.  In fact,
with respect to the diluted case, the latter diminishes by less than
$5\%$ up to $\rho \sigma^3=0.2$ and by less than $10\%$ up to $\rho \sigma^3=0.4$.
We note that simple dimensional considerations would
have suggested, instead, a proportional dependence of the surface
accessible area on $\langle R_g^2\rangle$.  The failure of the simple
dimensional analysis suggests that, despite ring sizes that are, on
average, smaller at higher densities, the corresponding configurations
are still sufficiently exposed to the solvent and maintain around each bead enough ``free room''
to accommodate the probe sphere used to measure the accessible surface.

{\it Geometry of ring portions.}
The results presented so far address properties of entire rings.
We shall next examine how various
ring portions, or arcs, are affected by intra-chain and
inter-chain entanglement. We first report on how the
mean square end-to-end distance, $R^2_{ee}(\ell)$ of arcs of contour
length $\ell$, depends on the solution density and ring topology.
Furthermore, for rings with non-trivial topology we shall
identify the smallest arc accommodating the knot and
examine how its contour length depends on the monomer density.

\begin{figure}
(a)\includegraphics[width=3.0in]{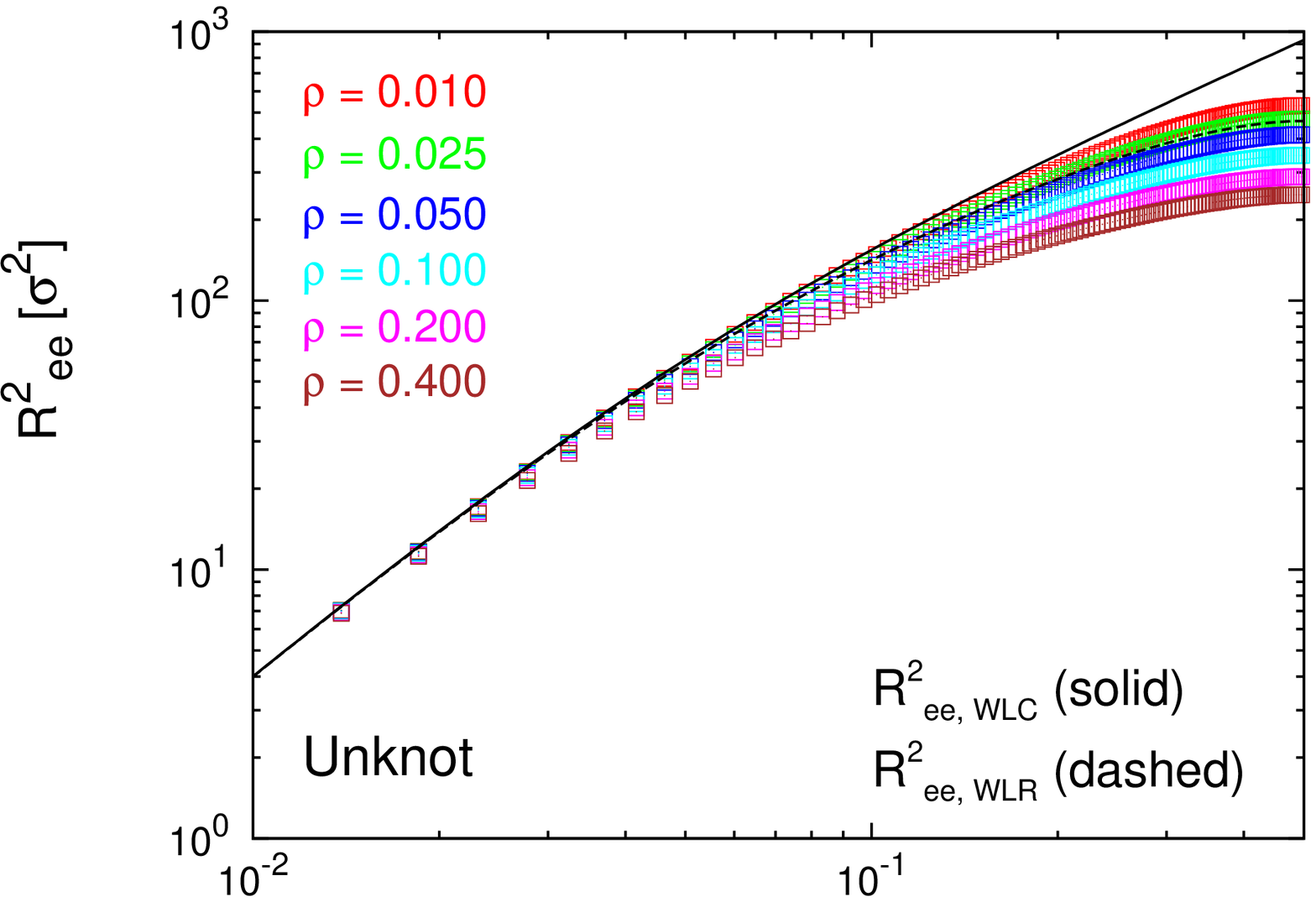}
(b)\includegraphics[width=3.0in]{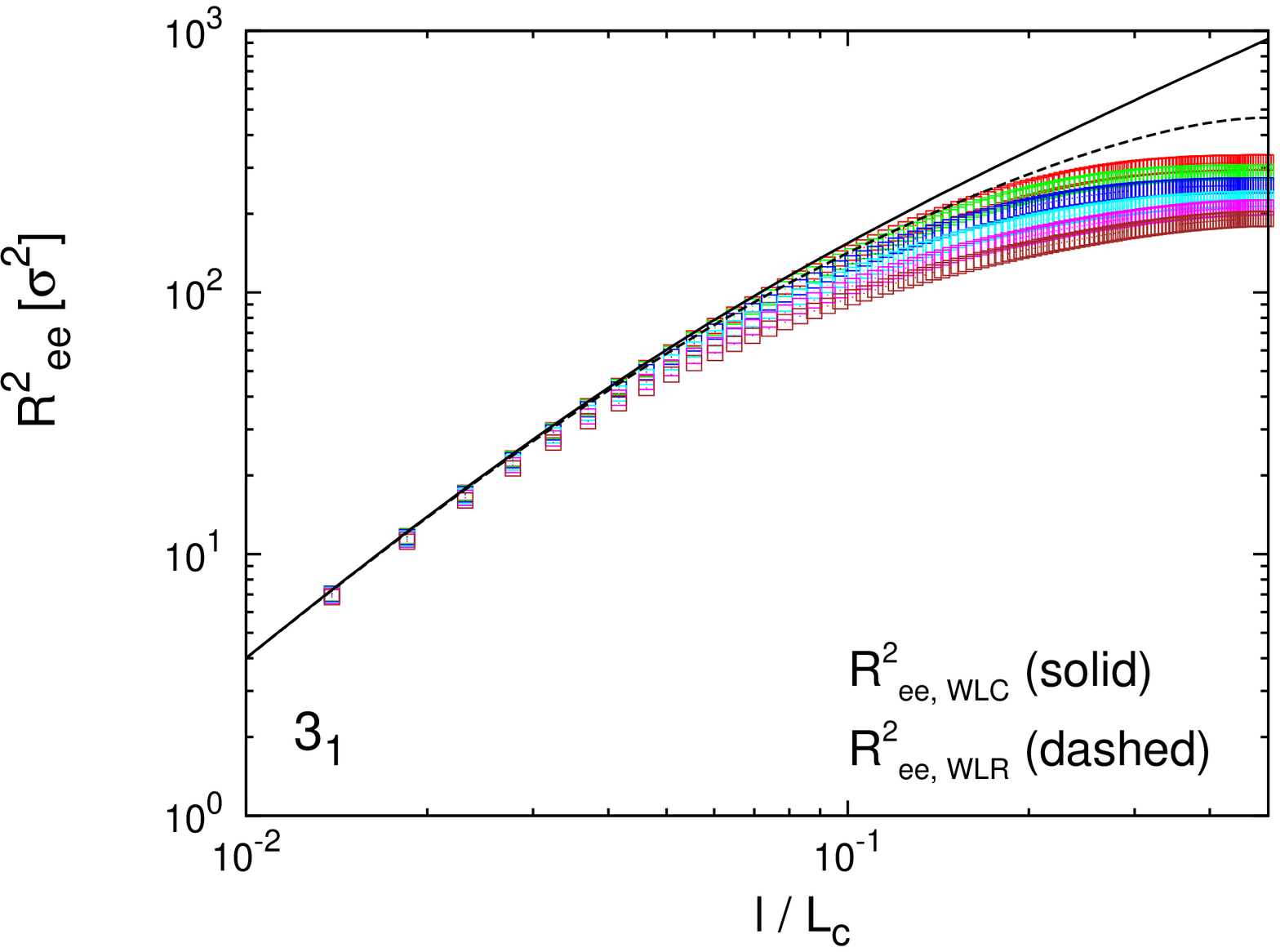}
\caption{\label{fig:r2vsn}
Average square spatial distance $R^2_{ee}$ between ring monomers
as a function of their (normalized) contour length separation, $\ell / L_c$, along the chain:
results for (a) unknotted and (b) trefoil-knotted ring polymers at increasing monomer densities.
}
\end{figure}

The density-dependent behavior of $R_{ee}^2(\ell)$ is shown in Fig. \ref{fig:r2vsn}.
The data for knotted and unknotted rings are presented in two separate panels.
For reference, in each panel we include the graph for the mean square end-to-end
distance for a worm-like chain (WLC):
\begin{equation}\label{eq:wlc}
R^2_{ee,WLC}(\ell) = \frac{\ell_K^2}{2} \left[ \frac{2 \ell}{\ell_K} + \exp\left(-\frac{2\ell}{\ell_K}\right) - 1 \right] .
\end{equation}
The analogous expression for a worm-like ring (WLR) is closely approximated by:
\begin{equation}\label{eq:wlr}
R^2_{ee,WLR}(\ell) = \left( \frac{1}{R^2_{ee,WLC}(\ell)} + \frac{1}{R^2_{ee,WLC}(L_c-\ell)} \right)^{-1} \ .
\end{equation}
The latter expression is obtained by matching analytically the
exact, small-$\ell$ (stiff)
and large-$\ell$ (flexible) limiting behaviors for a ring polymer.

Fig. \ref{fig:r2vsn} shows that for both unknotted and
trefoil-knotted rings at all densities, the WLC (and WLR) behavior is
followed closely only for arclengths smaller than about $\ell_K$ (corresponding to $\ell/L_c \sim 0.04$).
Beyond this arclength, the WLR behavior is followed well by
unknotted rings in dilute solutions. 
Noticeable differences from the WLR trend occur for both unknotted
and knotted rings at densities larger than $\rho^*\sigma^3=0.1$.
The deviations grow with:
(i) increasing solution density at fixed topology and arclength separation;
(ii) increasing arclength separation at fixed topology and density;
(iii) changing ring topology from unknotted to trefoil-knotted at fixed density and arclength separation. 
In summary, departures from the WLR trend are seen upon increasing
either the inter-chain (solution density) or the
intra-chain (topological) entanglements.

{\it Properties of the knotted portion of a ring.}
We conclude the analysis of the equilibrium properties by examining
how density influences the length of the ring portion
accommodating the knot, $L_{knot}$.  The analysis is motivated by
recent studies which showed that the typical value of $L_{knot}$ is
substantially larger for collapsed rings than for unconstrained or
swollen ones \cite{marcone2007,tubiana2,baiesi2011}.  In addition, for
knotted rings subject to spherical confinement, it was seen that the
increasing geometrical entanglement acquires a complex multiscale
character so that very different values of $L_{knot}$ are obtained
upon locating the knotted portion using bottom-up or top-down approaches (see Methods, Sec. \ref{sec:knotid}).
These properties are expected to significantly impact the physical behavior of the chains
(see e.g. Ref. \cite{micheletti_pnas} for a discussion of knot
localization effects in a biological context, namely the ejection of
tightly packed - and highly entangled - DNA from a viral capsid).

\begin{table}[htp]
\begin{center}
\begin{tabular}{|c|c|c|}
\hline
Monomer density, $\rho\sigma^{3}$  &  $ \langle L_{knot} \rangle $  &  $ \langle L_{knot} \rangle / L_c $\\
\hline
0.010  &  $ 88.8 \pm 0.2 $  &  $ 0.41 \pm 0.0010  $\\
\hline
0.025  &  $ 90.2 \pm 0.3 $  &  $ 0.42 \pm 0.0010  $\\
\hline                                                                                                   
0.050  &  $ 89.7 \pm 0.1 $  &  $ 0.42 \pm 0.0005 $\\
\hline
0.100  &  $ 85.9 \pm 0.1 $  &  $ 0.40 \pm 0.0005 $\\
\hline 
0.200  &  $ 79.9 \pm 0.1 $  &  $ 0.37 \pm 0.0005 $\\
\hline    
   0.400  &  $ 72.9 \pm 0.1 $  &  $ 0.34 \pm 0.0005 $\\
\hline
\end{tabular}
\caption{\label{tab:klensize}
Average size of the knotted portion of $3_1$ ring polymers, $\langle L_{knot} \rangle$
as a function of monomer density, $\rho$.
On average, the knot occupies a large fraction of the polymer chain.
}
\end{center}
\vspace{-0.6cm}
\end{table}

In the present context, the average value of $L_{knot}$ remains fairly
constant at all densities, see Table \ref{tab:klensize}. However, it
should be noted that above the overlap density, $\rho^*$, the
probability distribution function for $L_{knot}$ acquires a rounder
and broader shape, see Fig. \ref{fig:klen.eps}.
The results are robust across different methods used for locating the knot (see
Supporting Fig. 1 
for a comparison of top-down and bottom-up approaches).
In conclusion, the properties of knots in dense solutions differ qualitatively from the two most
typical conditions (ring collapse or ring spatial confinement) leading
to high monomer density.

\begin{figure}[htp]
\includegraphics[width=3.0in]{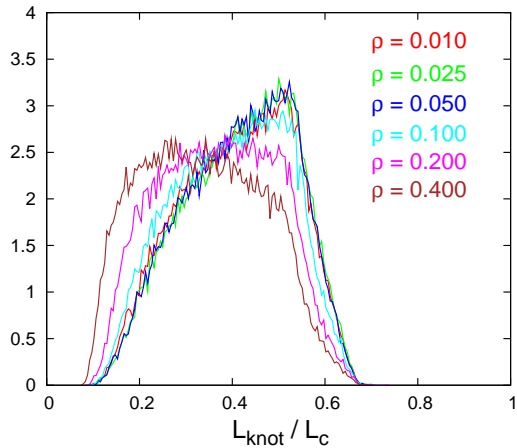}
\caption{
Probability distribution function of the normalized knot length, $L_{knot} / L_c$,
at different monomer concentrations.
\label{fig:klenpdf}
}
\end{figure}

\subsection{Dynamical properties}\label{sec:dynam}

The growing level of entanglement found in solutions of increasing
density is expected to strongly impact the ring dynamics.
By analogy with the case of linear polymers melts
\cite{doi,kremer_jcp,everaers_science,tzoumanekas2009onset,lahmar2009onset},
multiple dynamical regimes are expected.
For the latter system, it is known that at times longer
than the slowest chain relaxation time, the motion of the chain center of mass follows
the standard diffusion while at smaller times the kinetics is
dominated by the slow chain reptation inside the tube created by
inter-chain constraints \cite{kremer_jcp}.

We accordingly monitor several observables:
\begin{enumerate}
\item
The typical time scale for ring diffusion in space;
\item
The time scales dominating the fluctuations in rings size and orientation;
\item
Finally, for rings with non-trivial topology, we examine how the knotted portion moves
both in space and along the ring backbone.
\end{enumerate}
Some of the above mentioned quantities have been considered before
in studies on melts of unknotted rings
\cite{grosbergkremer,cates_ring1,cates_ring2,kremergrosberg2011_dyn},
in lattice models of isolated knotted rings
\cite{quake,lai,orlandinistella},
and in polymer models of knot diffusion along DNA \cite{bao_quake,voloknot}.
The present context therefore offers
an opportunity to examine the impact of both solution density and ring
topology on dynamics and compare the time-scales associated to the various phenomena.

{\it Time autocorrelation function for the radius of gyration.}
We start by considering the time autocorrelation function of an internal
(i.e. independent of the ring absolute space position and orientation)
quantity, namely the radius of gyration:
\begin{equation}
C_{R_g}(t) = \frac{ \langle R_g(t) \, R_g(0) \rangle - \langle R_g \rangle^2 } { \langle R_g^2 \rangle - \langle R_g \rangle^2 } .
\label{eq:rgautocorr}
\end{equation}
where the brackets, $\langle ... \rangle$, denote the average over
simulation time and over rings.  The behavior of $C_{R_g}(t)$ was
considered in recent lattice studies of {\it isolated}, unconstrained
knotted and unknotted rings \cite{lai,orlandinistella}.  These
investigations showed that $C_{R_g}$ decays with a characteristic
time, $\tau_{R_g}$, that is larger for knotted rings than for
unknotted ones \cite{lai,orlandinistella}.  The effect reflects the
enhanced self-hindrance of rings with non-trivial topology; consider,
for instance, that the minimal number of crossings observed in any
two-dimensional projection of an unknotted ring is zero, while it is
three for trefoil-knotted rings (which hence must necessarily wind on
themselves more than unknots).

Investigating the decay properties of $C_{Rg}(t)$ in the present
context serves a twofold purpose.  On the one hand, it can clarify
whether the above-mentioned lattice results apply to off-lattice
contexts, too.  On the other hand, by analyzing the dependence of
$C_{R_g}(t)$ on solution density, it is possible to assess if, and to
what extent, the increased intra-chain and inter-chain entanglement
affect differently the size relaxation times of unknotted and knotted
rings.

\begin{figure*}
$$
\begin{array}{lr}
(a)\includegraphics[width=3.0in]{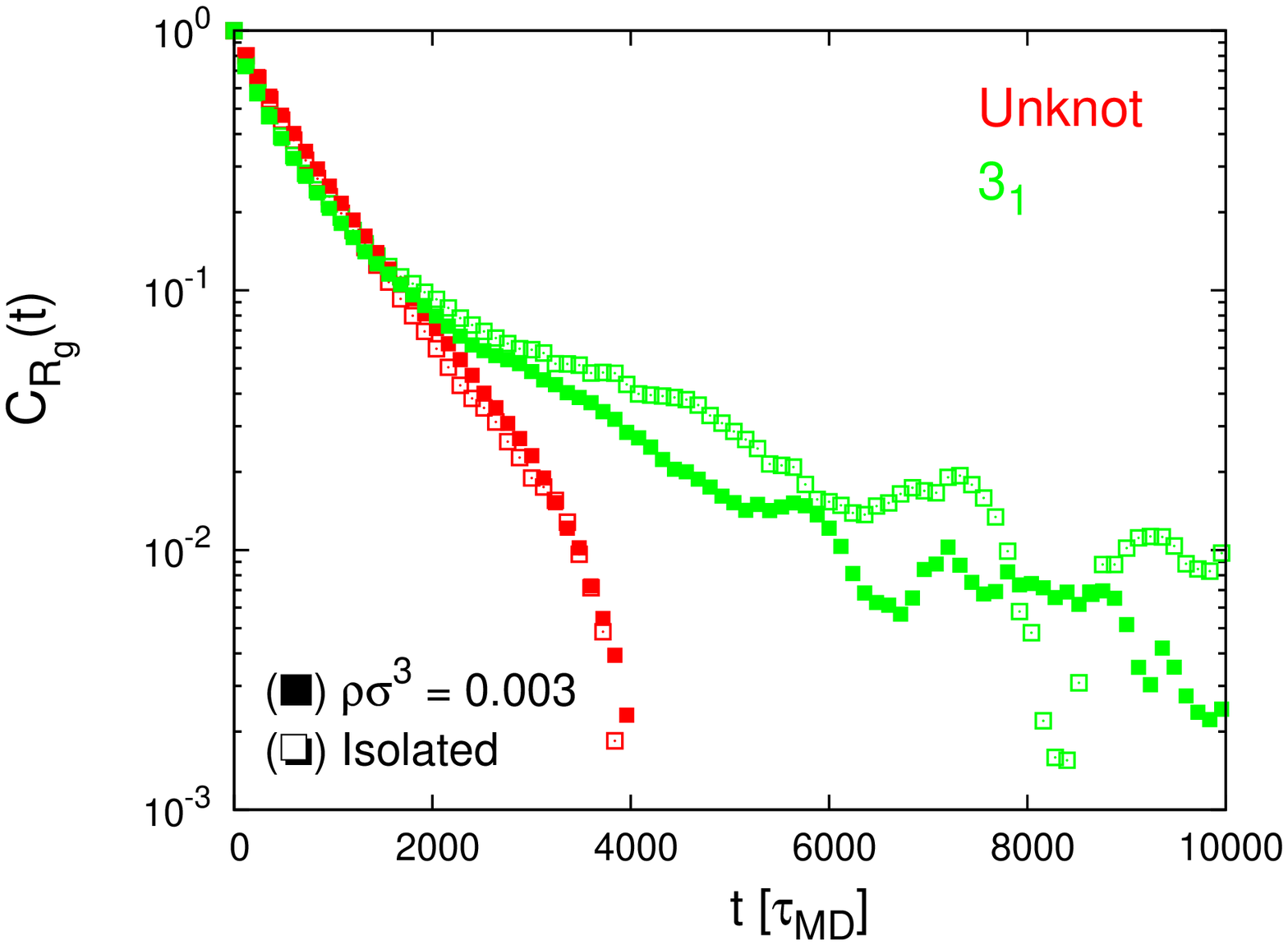} &
(b)\includegraphics[width=3.0in]{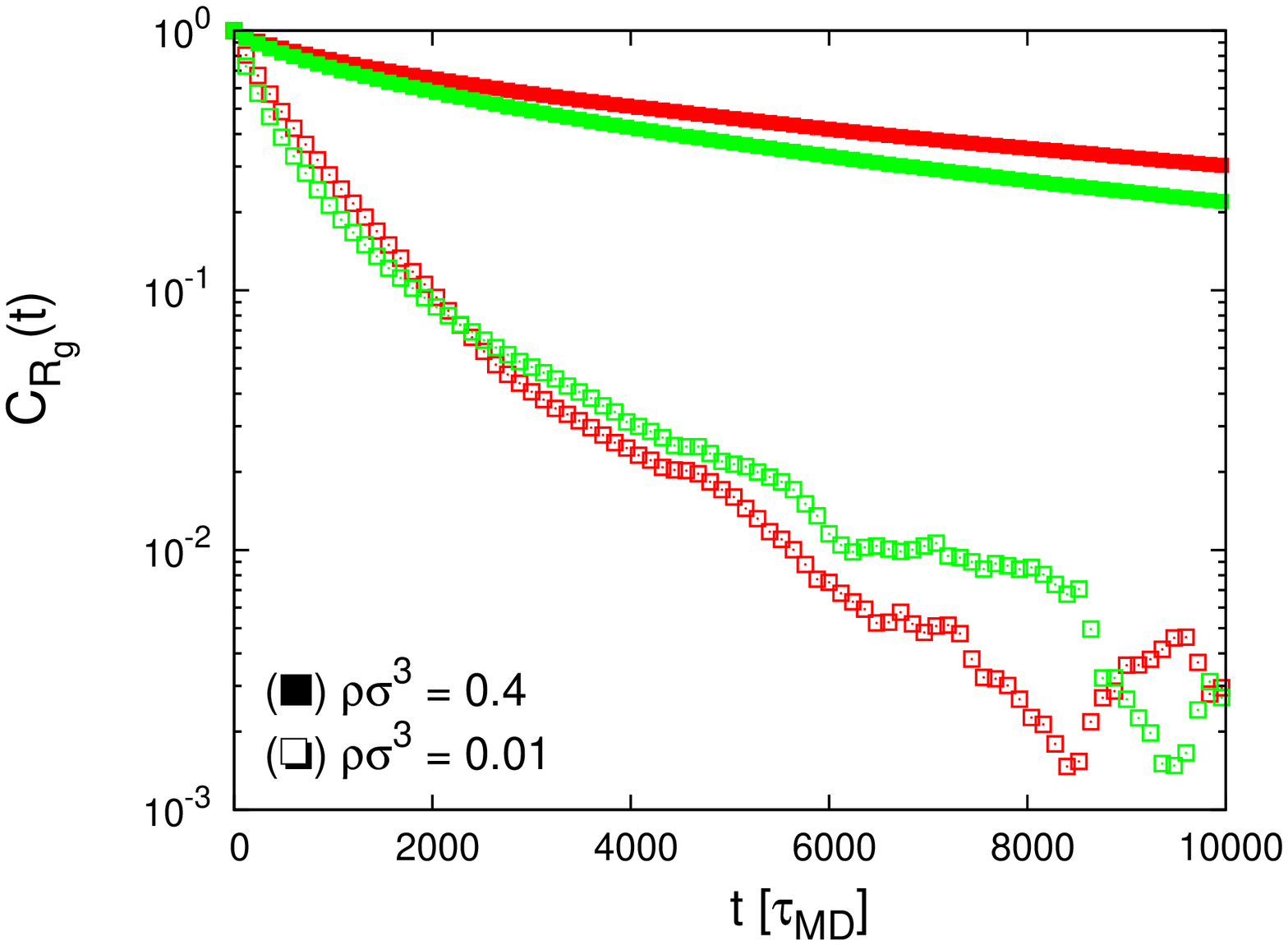}
\end{array}
$$
\caption{
\label{fig:reautocorr}
Radius of gyration time correlation function $C_{R_g}(t)$ (Eq. \ref{eq:rgautocorr})
for unknots (red symbols) and trefoils (green symbols).
(a):
Result for isolated rings (open symbols) and rings in very dilute conditions with $\rho \sigma^3 = 0.003$ (filled symbols).
The long time behavior of $C_{R_g}(t)$ for $3_1$ rings deviates from the $C_{R_g}(t)$ for the unknots,
and shows a longer exponential tail, in agreement
with previous studies of isolated ring polymers on lattice \cite{quake,lai,orlandinistella}.
(b):
Results for $\rho \sigma^3 = 0.01$ (open symbols) and $\rho \sigma^3 = 0.4$ (filled symbols).
}
\end{figure*}

According to this plan we first compute $C_{R_g}$ from simulations of
{\it isolated} knotted and unknotted rings. The results are shown in
Fig. \ref{fig:reautocorr}a (open symbols) and fully support the
lattice results regarding the slower relaxation of trefoils compared to unknots.
For rings in solutions, at all considered  densities ($\rho \sigma^3 = 0.01 - 0.4$),
one is surprised to find that the $C_{R_g}$ curves for trefoils and
unknots display much smaller differences compared
to the isolated case, see Fig. \ref{fig:reautocorr}b.  Indeed, a
behavior quantitatively similar to the isolated case is found only at
densities lower than $\rho \sigma^3 =0.003$, as shown in
Fig. \ref{fig:reautocorr}a. 
The effect is quantitatively captured by calculating the effective
relaxation time, $\tau_{R_g}$, given by the integral of $C_{R_g}(t)$,
see Table \ref{tab:tau_rg_diam}. The data in the table illustrate vividly that
the density-dependent increase of the intra- and inter-chain
entanglement obliterates differences in the average
kinetic behavior of rings with different topological state.
For both unknots and trefoils, $\tau_{R_g}$ has an order of magnitude
increase over the considered range of solution density, $\rho$. Notice
that upon increasing $\rho$, the value of $\tau_{R_g}$ for unknots
overtakes the one of trefoils (see also the order of the red and green
curves in panels a and b in Fig. \ref{fig:reautocorr}).

\begin{table*}[htp]
\begin{center}
\begin{tabular}{|c||c|c||c|c||c|c|}
\hline
{}  &  UN  &  $3_1$  &  UN  &  $3_1$  &  UN  &  $3_1$\\
Monomer density, $\rho\sigma^{3}$ &  $\tau_{R_g}$  &  $\tau_{R_g} $  &  $\tau_{diam} $  &  $\tau_{diam} $ &  $\tau_{CM}$  &  $\tau_{CM}$\\
\hline
Single ring  &  $ (6.6 \pm 0.6) \cdot 10^2 $  &  $ (7.9 \pm 0.6) \cdot 10^2 $  &  $ (2.8 \pm 0.1) \cdot 10^3 $  &  $ (1.6 \pm 0.1) \cdot 10^3 $  &  $ 3.1 \cdot 10^3 $  &  $ 2.2 \cdot 10^3 $\\
\hline
0.003  &  $ (7.0 \pm 0.6) \cdot 10^2 $  &  $ (7.0 \pm 0.6) \cdot 10^2 $  &  $ (2.7 \pm 0.1) \cdot 10^3 $  &  $ (1.6 \pm 0.1) \cdot 10^3 $  &  $ 3.4 \cdot 10^3 $  &  $ 2.4 \cdot 10^3  $\\
\hline
0.010  &  $ (8.1 \pm 0.6) \cdot 10^2 $  &  $ (7.3 \pm 0.6) \cdot 10^2 $  &  $ (2.7 \pm 0.4) \cdot 10^3 $  &  $ (1.6 \pm 0.3) \cdot 10^3 $  &  $ 4.8 \cdot 10^3 $  &  $ 3.6 \cdot 10^3 $\\ 
\hline                                                                                                                                       
0.025  &  $ (1.1 \pm 0.1) \cdot 10^3 $  &  $ (7.8 \pm 0.6) \cdot 10^2 $  &  $ (2.8 \pm 0.4) \cdot 10^3 $  &  $ (1.6 \pm 0.3) \cdot 10^3 $  &  $ 4.8 \cdot 10^3 $  &  $ 3.6 \cdot 10^3 $\\ 
\hline                                                                                                                                        
0.050  &  $ (1.6 \pm 0.1) \cdot 10^3 $  &  $ (1.0 \pm 0.1) \cdot 10^3 $  &  $ (3.1 \pm 0.4) \cdot 10^3 $  &  $ (1.7 \pm 0.3) \cdot 10^3 $  &  $ 6.0 \cdot 10^3 $  &  $ 3.6 \cdot 10^3 $\\
\hline                                                                                                                                        
0.100  &  $ (2.7 \pm 0.1) \cdot 10^3 $  &  $ (1.5 \pm 0.1) \cdot 10^3 $  &  $ (3.9 \pm 0.4) \cdot 10^3 $  &  $ (2.2 \pm 0.3) \cdot 10^3 $  &  $ 7.2 \cdot 10^3 $  &  $ 4.8 \cdot 10^3 $\\
\hline                                                                                                                                        
0.200  &  $ (4.7 \pm 0.1) \cdot 10^3 $  &  $ (2.6 \pm 0.1) \cdot 10^3 $  &  $ (6.1 \pm 0.4) \cdot 10^3 $  &  $ (3.3 \pm 0.3) \cdot 10^3 $  &  $ 1.2 \cdot 10^4 $  &  $ 7.2 \cdot 10^3 $\\
\hline                                                                                                                                       
0.400  &  $ (1.25 \pm 0.01) \cdot 10^4 $  &  $ (7.3 \pm 0.1) \cdot 10^3 $  &  $ (1.59 \pm 0.04) \cdot 10^4 $  &  $ (9.2 \pm 0.4) \cdot 10^3 $  &  $ 3.9 \cdot 10^4 $  &  $ 2.2 \cdot 10^4 $\\
\hline
\end{tabular}
\caption{\label{tab:tau_rg_diam} 
Correlation times $\tau_{R_g}$ and $\tau_{diam}$,
calculated by numerical integration of the respective time correlation functions
$C_{R_g}(t)$ (Eq. \ref{eq:rgautocorr}) and $C_{diam}(t)$ (Eq. \ref{eq:e2ecf}).
Numerical integration is limited to the time interval where correlation functions are $> 10^{-2}$.
$\tau_{CM}$ is defined as the typical time required by rings
to transverse a region of linear size comparable to the ring
average gyration radius, i.e. $\delta r_{CM}^2(\tau_{CM}) = \langle R_g^2 \rangle$ (see Eq. \ref{eq:ringcommsd}).
All times are expressed in units of $\tau_{MD}$.
}
\end{center}
\vspace{-0.6cm}
\end{table*}

{\it Reorientation time.}
We now turn to consider kinetic properties that do depend on the absolute orientation of rings in space.
For linear polymers, it is customary to consider the autocorrelation
function of the end-to-end vector \cite{doi} which, for closed chains,
admits several generalizations \cite{kremergrosberg2011_dyn}.
The one considered here is the time correlation of the ring diameter vector:

\begin{equation}\label{eq:e2ecf}
C_{diam}(t) =
\frac{ \langle \vec d(t) \cdot \vec d(0) \rangle } { \langle | \vec d |^2 \rangle } ,
\end{equation}

\noindent where $\vec d$ is the vector joining two monomers that are
``diametrically opposite'' on the ring backbone, i.e. monomers with
the largest possible chemical distance, $L_c/2$.
Here the brackets $\langle \cdots \rangle$ denote a multiple average:
over the dynamical trajectory and over each pair of diametrically opposite beads for
every ring in solution.
Notice that $C_{diam}$ is sensitive to both changes in modulus
of the diameter vector and to its absolute orientation.
For this reason, $\tau_{diam}$, the characteristic decay time of
$C_{diam}(t)$ is customarily referred to as the {\it reorientation} time.

\begin{figure*}
$$
\begin{array}{lr}
(a)\includegraphics[width=3.0in]{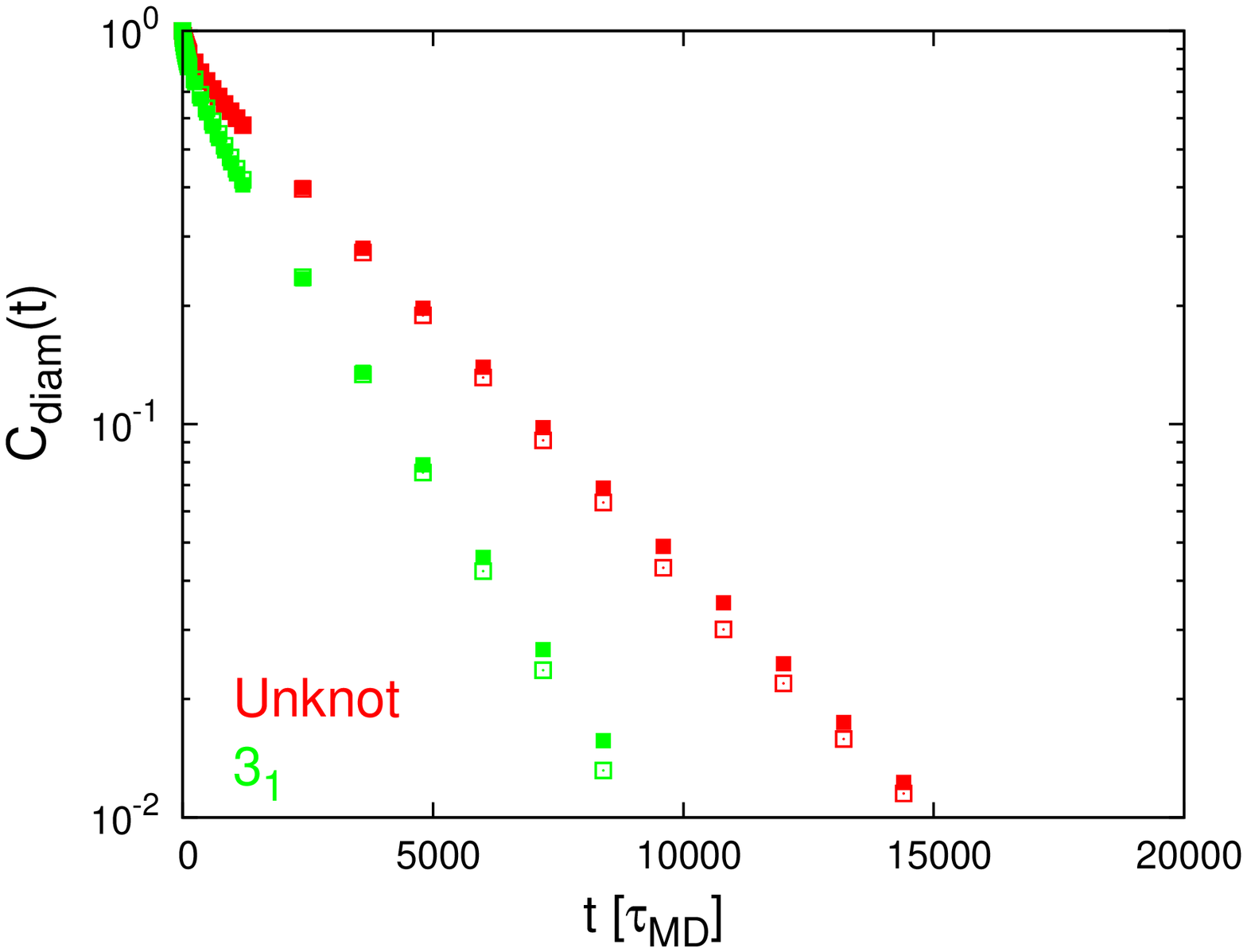} &
(b)\includegraphics[width=3.0in]{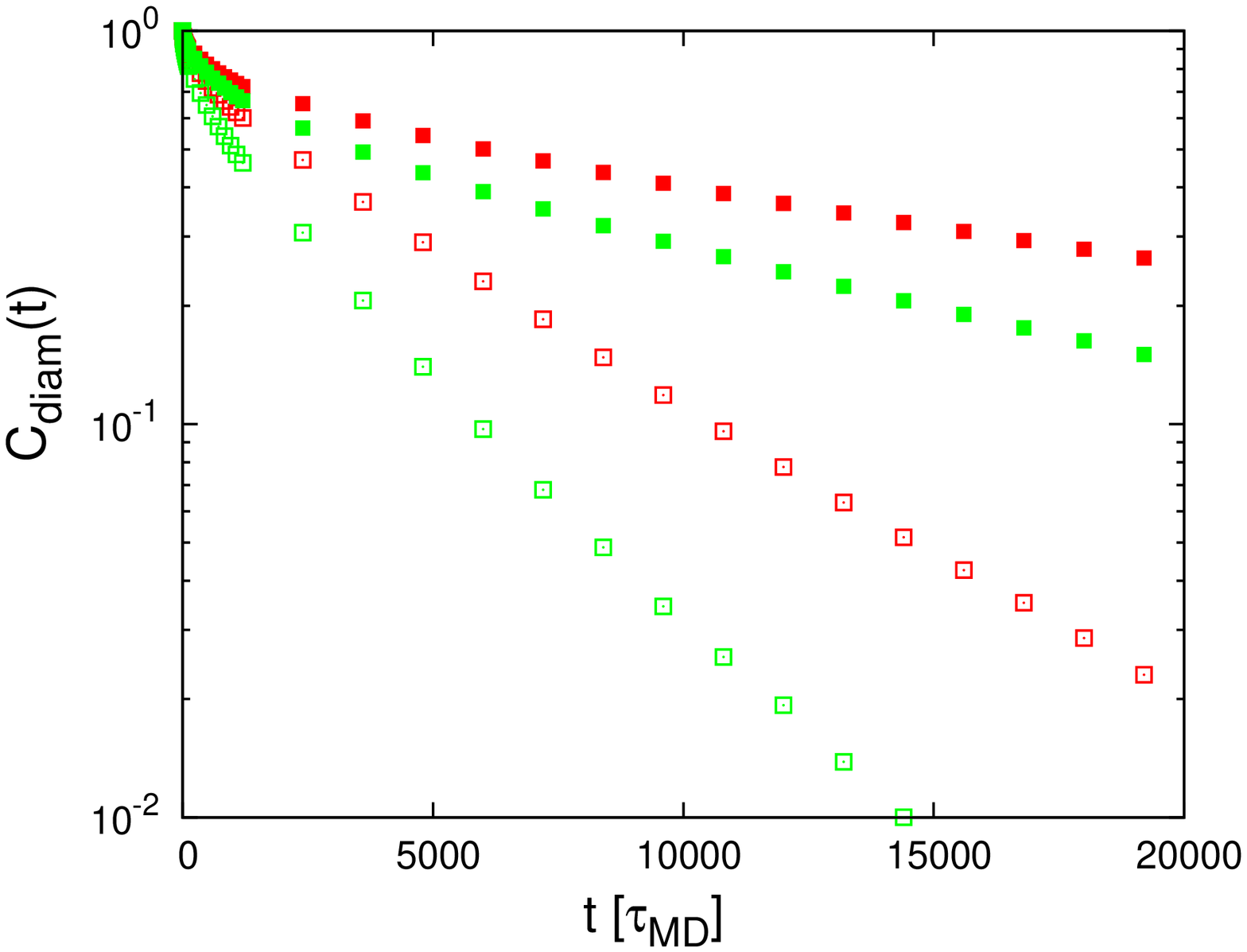}
\end{array}
$$
\caption{
\label{fig:e2ecf}
Ring diameter time correlation function $C_{diam}(t)$ (Eq. \ref{eq:e2ecf})
for the unknots (red symbols) and trefoils (green symbols).
(a):
$\rho \sigma^3 = 0.01$ (open symbols) and $\rho \sigma^3 = 0.025$ (filled symbols).
(b):
$\rho \sigma^3 = 0.1$ (open symbols) and $\rho \sigma^3 = 0.4$ (filled symbols).
}
\end{figure*}

Fig. \ref{fig:e2ecf} portrays $C_{diam}(t)$ for unknots (red symbols) and $3_1$ knots (green symbols)
at different monomer concentrations.
Analogously to $\tau_{R_g}$, we define the corresponding $\tau_{diam}$
as the numerical integral of $C_{diam}(t)$.
Final results are reported in Table \ref{tab:tau_rg_diam}.
As for $C_{R_g}(t)$, the asymptotic decay time of $C_{diam}(t)$ depends on ring topology
and is faster for trefoils than unknots.
To the best of our knowledge this effect, that holds at all densities, was neither pointed out nor addressed before.
Extending this analysis and considerations to rings with more complicated knot types
could represent an interesting avenue for further investigations.

\begin{figure*}
(a)
\includegraphics[width=3.0in]{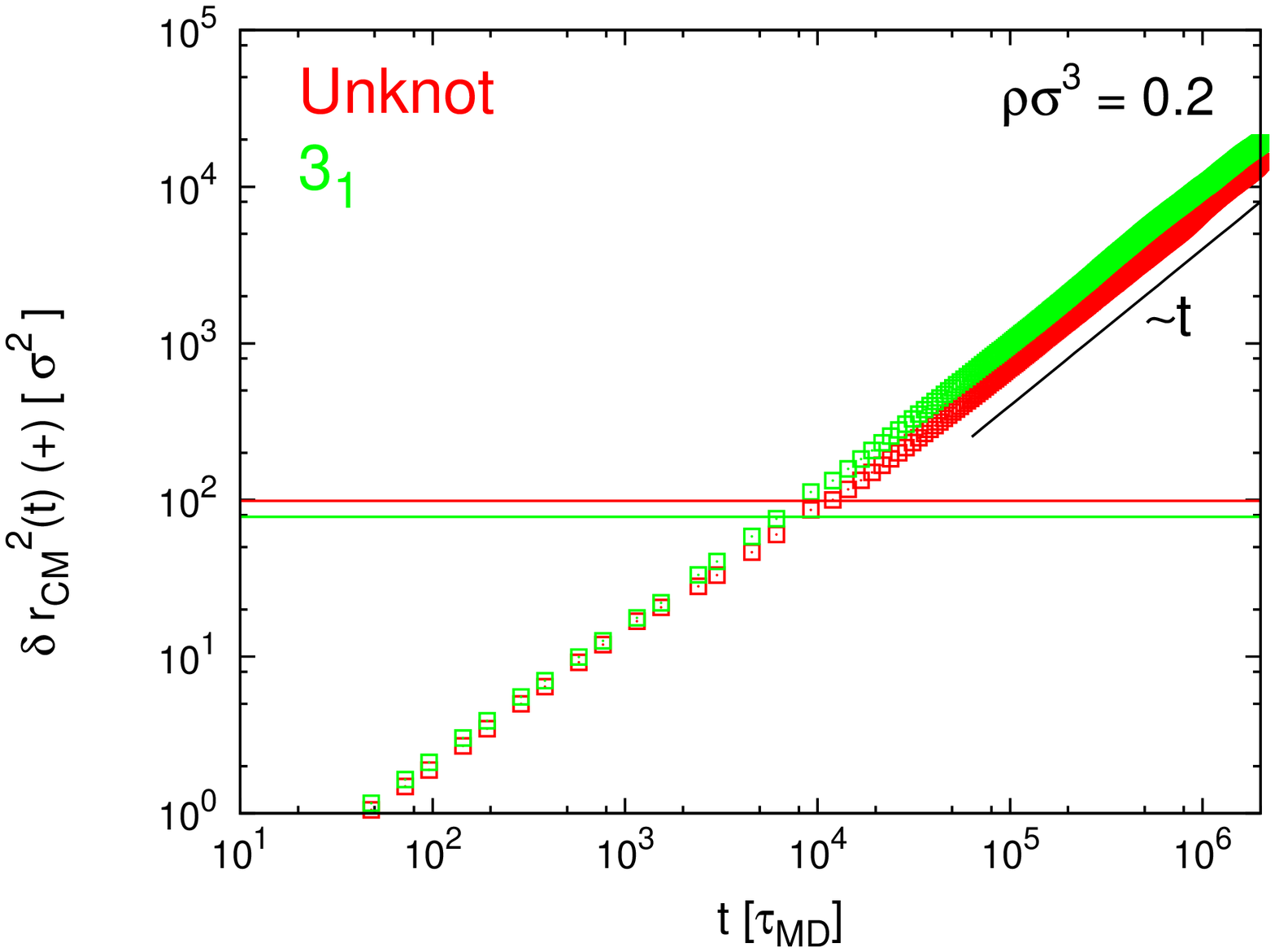}
(b)
\includegraphics[width=1.7in]{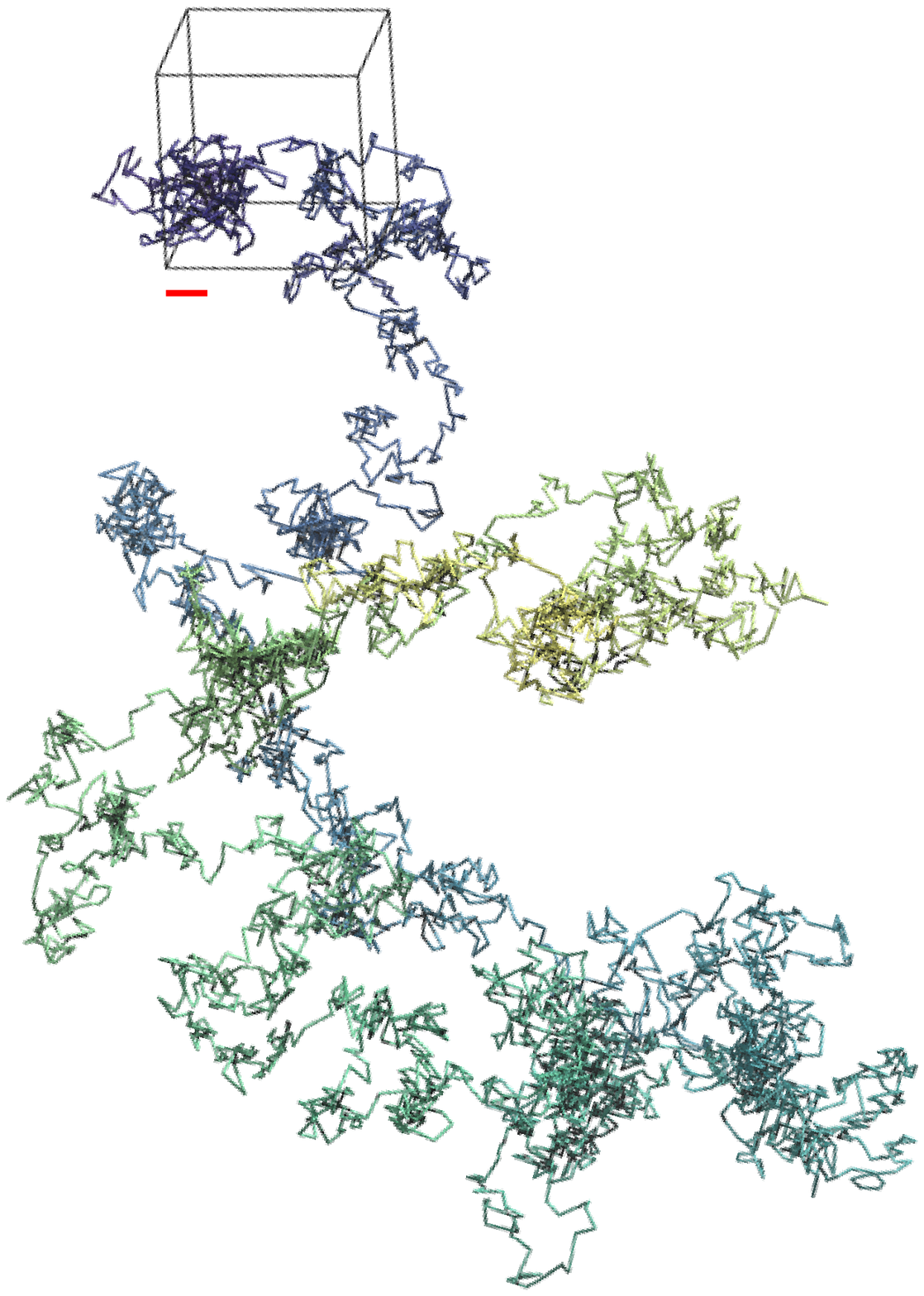}
\caption{
\label{fig:com_msd}
(a):
Comparison between the time mean-square displacement of the rings
center of mass ($\delta r_{rCM}^2(t)$, Eq. \ref{eq:ringcommsd}) for
unknots (red symbols) and trefoils (green symbols) and
the corresponding average square gyration radii (horizontal lines), for $\rho \sigma^3 = 0.2$.
Within the considered time-span, rings diffuse over distances much larger than their typical sizes.
(b):
Diffusive motion of the center of mass of a randomly selected trefoil, for $\rho \sigma^3 = 0.2$.
The red bar denotes the average ring size, $\langle R_g^2 \rangle^{1/2}$.
Corresponding plots at different monomer densities look qualitatively similar.
}
\end{figure*}

{\it Diffusion of the ring center of mass.}
Because trefoils are more compact than unknots they
should diffuse faster in the solution, analogously to what they do in other
types of media \cite{gonzalez2004dynamics,weber2006numerical}.
This intuitive expectation is indeed confirmed by inspecting Fig. \ref{fig:com_msd},
which portrays the mean square displacement of
the ring center of mass, $\delta r_{CM}^2 (t)$,
for time lags of increasing duration, $t$:
\begin{equation}\label{eq:ringcommsd}
\delta r_{CM}^2(t) \equiv \langle \left( {\vec r}_{CM} (t) - {\vec r}_{CM} (0) \right)^2 \rangle .
\end{equation}
The displacement data given in Fig. \ref{fig:com_msd} pertain to
$\rho\sigma^3 = 0.2$; similar plots are obtained for different values
of $\rho$.  It is seen that over the time-span covered by the
simulations, both unknotted and trefoil rings diffuse over distances
that exceed by several orders of magnitude their average size.  For
traveled distances larger than the average ring size, the motion of
the center of mass follows standard diffusion.
The corresponding diffusion time $\tau_{CM}$,
defined as $\delta r_{CM}^2 (\tau_{CM}) = \langle R_g^2 \rangle$,
is given in Table \ref{tab:tau_rg_diam}.
Notice, that the values of $\tau_{CM}$ are comparable to the values of
$\tau_{diam}$ at the same solution density, consistently with
the case of linear polymers \cite{doi}.

{\it Backbone motion of the knot.} 
Finally, we examine the dynamics in space and along the ring backbone of the knotted portion of the
trefoils.  This is a computationally-demanding task, because the
numerically-costly identification of the knotted portion must be
carried out for each ring of all sampled system snapshots.

\begin{figure*}
(a)
\includegraphics[width=2.9in]{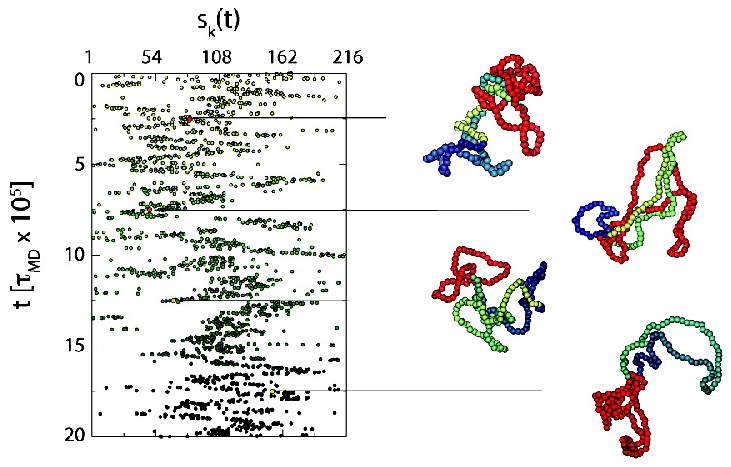}
(b)
\includegraphics[width=2.9in]{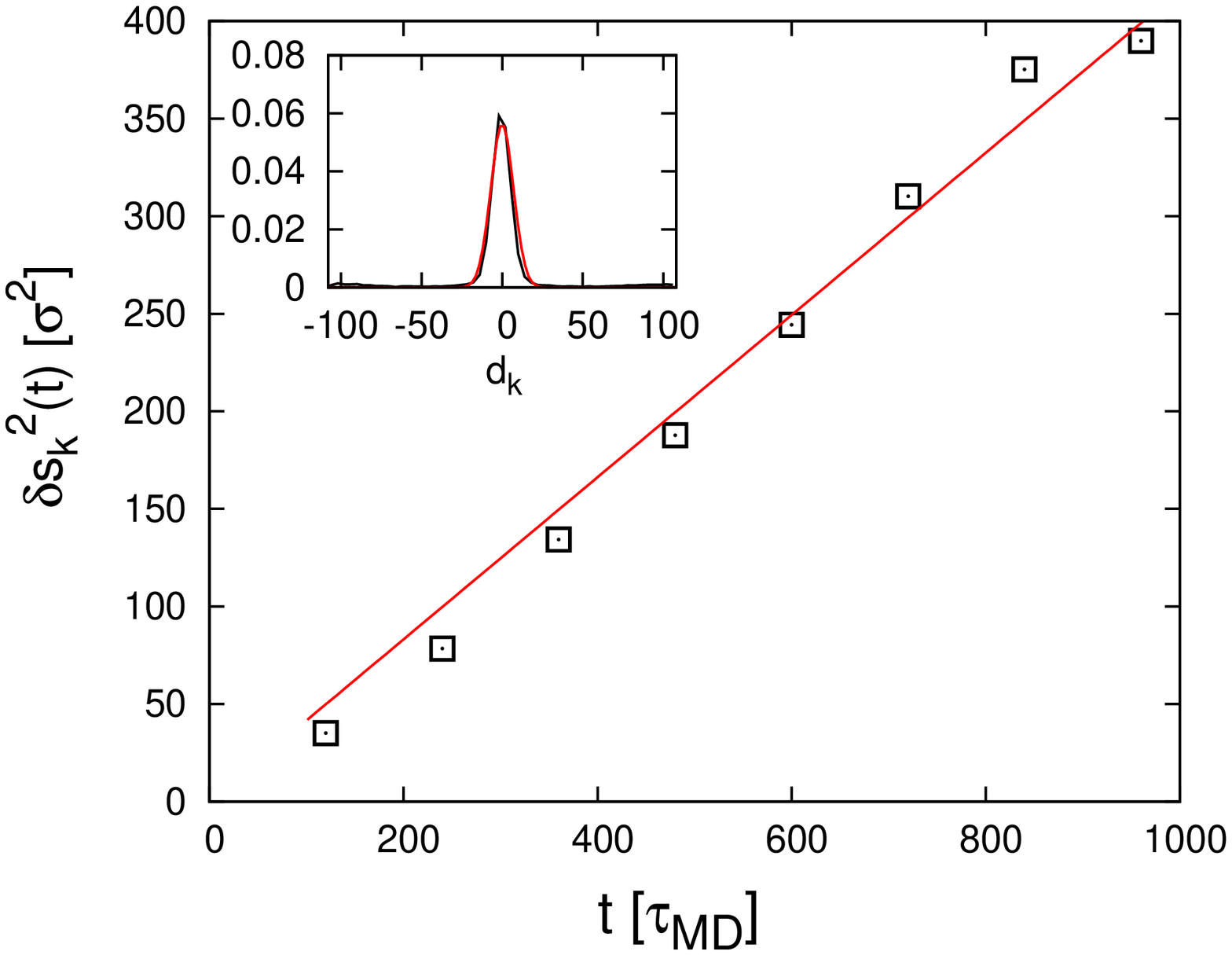}
\caption{
\label{fig:kjumpscal}
(a)
Time behavior of the knot midpoint position along the ring contour.
Midpoint coordinates are defined with respect to a reference monomer chosen at the beginning of the simulation.
On the right, four ring configurations sampled at different times along the trajectory,
with the corresponding knotted portions marked in red.
(b)
Mean square displacement of the knot along the ring contour, $\delta s_k^2(t)$ (Eq. \ref{eq:dsk2}),
at density $\rho \sigma^3 = 0.1$.
The proportional dependence on time (red line passing through the origin) can be expressed as $2 D_k t$,
where $D_k$ is the one-dimensional diffusion coefficient:
for the case shown, $D_k = 0.21 \sigma^2 / \tau_{MD}$
(see Table \ref{tab:knotdiff}, for a list of diffusion coefficients at all monomer concentrations).
{\it Inset}:
The Gaussian distribution (red curve) associated with the model one-dimensional diffusion process matches well
the numerical distribution (black line) of incremental knot displacements, $d_k$ (Eq. \ref{knotdispl}).
}
\end{figure*}

For simplicity, the instantaneous position of the knot on the ring is
taken to coincide with the chemical coordinate of the midpoint of the
knotted portion (defined with respect to the absolute monomer indexing
of the ring set at the beginning of the simulation).
The instantaneous knot midpoint position, $s_k(t)$ is recorded at fixed time intervals, $\Delta t = 120 \tau_{MD}$.
The incremental knot displacement between time snapshots $n \left(= \frac{t}{\Delta t}\right)$ and $n+1$,
$d_k(n)$, is computed using the expression:
\begin{equation}\label{knotdispl}
d_k(n) = \left\{ \left[ s_k(t+\Delta t) - s_k(t) + \frac{3}{2} L_c \right] \mbox{ mod } L_c \right \} - \frac{L_c}{2}
\end{equation}
where ``mod'' denotes the modulus operation.
The previous expression ensures that the displacement is mapped in the $[-L_c/2, L_c/2]$ interval.
The value of $\Delta t$ is sufficiently small that,
at all considered densities, the occurrence of large ``jumps'' (say
equal or larger than $L_c/4$ in modulus) is rare.
This is illustrated in the inset of Fig. \ref{fig:kjumpscal} (for $\rho \sigma^3 = 0.1$) which
further highlights the Gaussian character of the distribution.

The knot mean square displacement on the ring contour,
$\delta s_k^2(t)$, is defined by
\begin{equation}\label{eq:dsk2}
\delta s_k^2(t = i \cdot \Delta t) = \left \langle \left( \sum_{j=0}^{i-1} d_k(j) \right)^2 \right \rangle ,
\end{equation}
where the summation index runs over consecutive displacements,
and the average $\langle ... \rangle$ is taken
(1) over the trajectory and (2) over the rings ensemble.
Fig. \ref{fig:kjumpscal} illustrates that a
linear relationship holds between $\delta s_k^2(t)$ and time,
indicating that the knot sliding on the ring backbone can be
treated as a one-dimensional diffusion process.
The corresponding diffusion coefficients, $D_k$, calculated at different solution
densities are given in Table \ref{tab:knotdiff} which also reports the typical time, $\tau_k$, required by the knot to
diffuse by its average length on the ring backbone.
Interestingly, $\tau_k$ is almost unaffected by increasing monomer concentration.
In addition, for a given solution density, $\tau_k$ exceeds all other characteristic times  given in Table \ref{tab:tau_rg_diam}.

\begin{table}[h]
\begin{center}
\begin{tabular}{|c|c|c|}
\hline
Monomer density, $\rho$ & $D_k [\sigma^2 / \tau_{MD}]$ & $\tau_k [\tau_{MD}]$\\
\hline
0.010 &  $0.18 \pm 0.01$  &  $2.1 \cdot 10^4$ \\
\hline
0.025 &  $0.19 \pm 0.01$  &  $2.0 \cdot 10^4$ \\
\hline
0.050 &  $0.19 \pm 0.01$  &  $2.0 \cdot 10^4$ \\
\hline
0.100 &  $0.21 \pm 0.01$  &  $1.8 \cdot 10^4$ \\
\hline
0.200 &  $0.23 \pm 0.01$  &  $1.6 \cdot 10^4$ \\
\hline
0.400 &  $0.16 \pm 0.02$  &  $2.3 \cdot 10^4$ \\
\hline
\end{tabular}
\caption{\label{tab:knotdiff} Diffusion coefficients, $D_k$ for the
  motion of the knotted portion along the ring contour length. The
  value of $D_k$ corresponds to half the slope of the line that,
  passing through the origin, provides the best interpolation to
  $\delta s_k^2(t)$ vs $t$.  The typical diffusion (sliding) time,
  $\tau_k$, is defined as the time required to diffuse by a backbone
  distance equal to the typical knot contour length ($\approx 40\%$ of
  the ring contour length, see Table \ref{tab:klensize}).  }
\end{center}
\vspace{-0.6cm}
\end{table}

{\it Motion of the knotted portion in space.}  The above observation
indicates that the knot location on the ring backbone is practically
quenched as a ring diffuses in space over a distance equal to its size.
This suggests that the three-dimensional motion of the knotted portion
should be akin to the one of a fixed, tagged, portion of the ring
of comparable size.

\begin{figure*}
$$
\begin{array}{cc}
(a)\includegraphics[width=3.0in]{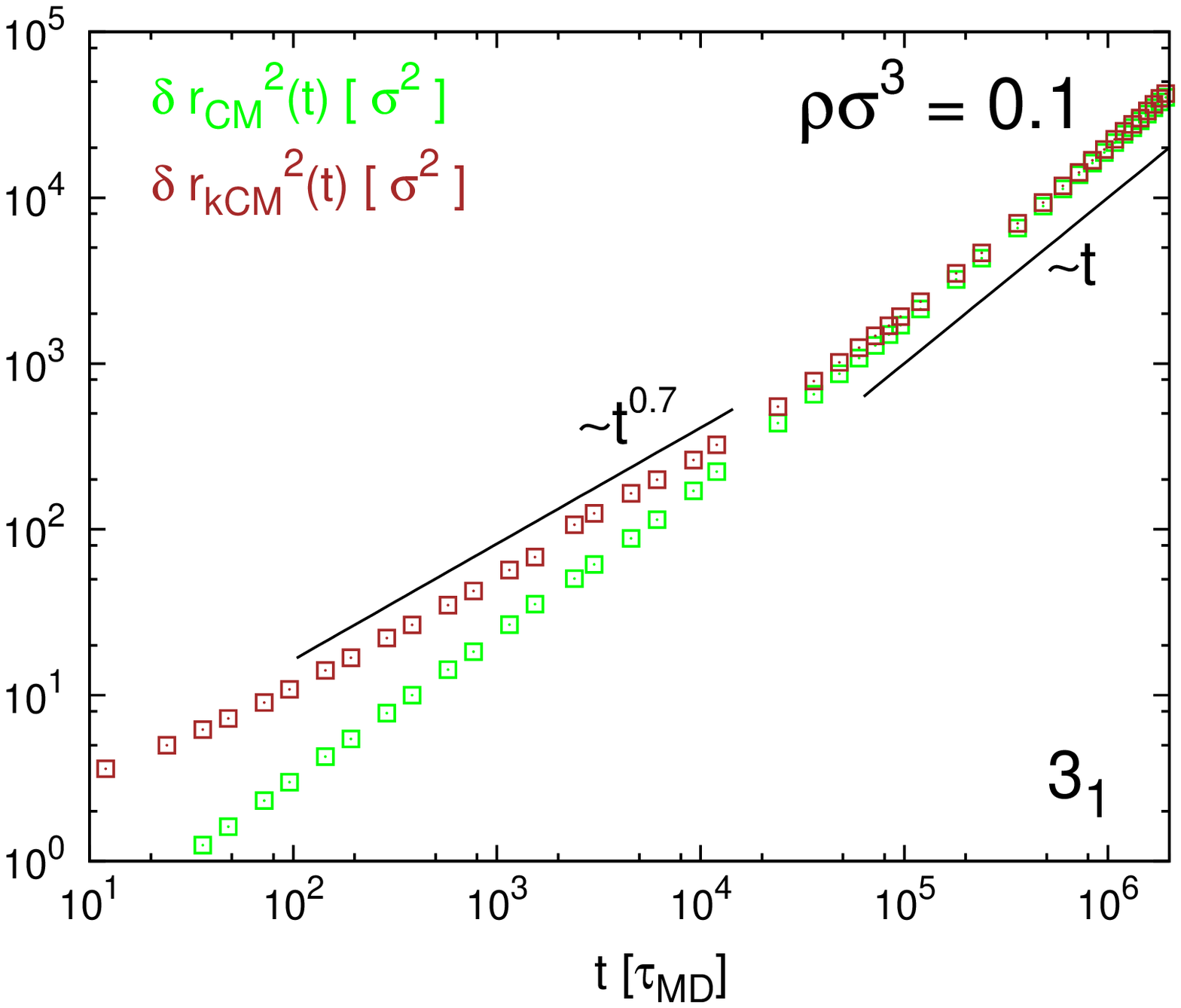} &
(b)\includegraphics[width=3.0in]{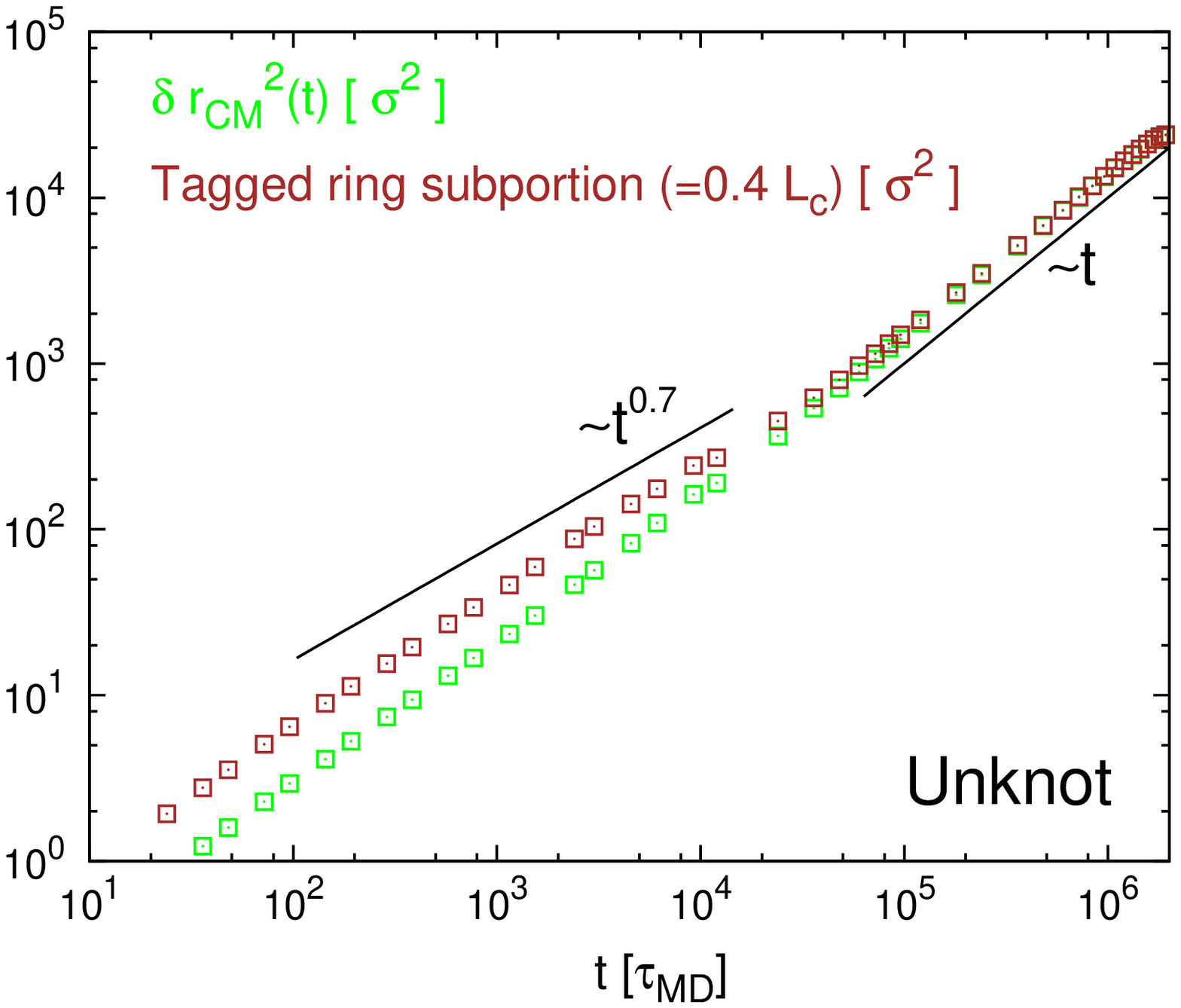}
\end{array}
$$
\caption{
\label{fig:knot_msd}
(a)
Comparison between the mean-square displacement of the center of mass of trefoils
($\delta r_{CM}^2(t)$, Eq. \ref{eq:ringcommsd}, green symbols)
and the center of mass of their knotted portion ($\delta r_{kCM}^2(t)$, Eq. \ref{eq:dsknot}, brown symbols).
The motion of the knotted portion is sub-diffusive at short times, and follows the chain global displacement at larger times.
(b)
Comparison between the mean-square displacement of the center of mass of unknots
($\delta r_{CM}^2(t)$, Eq. \ref{eq:ringcommsd}, green symbols)
and the center of mass of a tagged portion (brown symbols) of linear size
equal to the average knot length on $3_1$ rings 
($\approx 0.4 L_c$, Table \ref{tab:klensize}).
}
\end{figure*}

This observation was verified by considering  the time dependence of the 
mean-square displacement of the knot center of mass,  defined 
(analogously to Eq. \ref{eq:ringcommsd}) as:
\begin{equation}\label{eq:dsknot}
\delta r_{kCM}^2(t) = \langle \left( {\vec r}_{kCM} (t) - {\vec r}_{kCM} (0) \right)^2 \rangle .
\end{equation}
Fig. \ref{fig:knot_msd} (left panel) shows that the diffusive behavior of the 
observable ${\vec r}_{kCM}$ sets in at times larger than the relaxation time of the 
whole chain (compare to Fig. \ref{fig:com_msd}),
while a subdiffusive behavior $\sim t^{\alpha}$ with $\alpha \approx 0.7$ is seen at smaller times.
Notice that the same exponent, $\alpha \approx 0.7$ was previously observed
by looking at the stochastic dynamics of the center of mass of the knotted portion
in isolated self-avoiding polygons on a cubic lattice \cite{orlandinistella}.
Analogous plots at different monomer densities show the same subdiffusive behavior 
with an exponent $\alpha$ that decreases slightly with $\rho$ (not shown). 
We have then calculated the mean square displacement of a
randomly-picked ring portion spanning $40 \%$ of the contour length
(the typical size of the knotted portion, Table \ref{tab:klensize})
of the unknotted ring. As shown in Fig. \ref{fig:knot_msd}b it was found 
that the tagged portion moves very compatibly with the motion of the knotted part,
showing an analogous crossover from a diffusive to a subdiffusive behavior.
This suggests that the underlying mechanism of the subdiffusive regime of the ``quenched'' knotted part 
is similar to the one that governs the subdiffusive behavior of any other (topologically 
trivial) tagged subregion of the ring.
Note, however, that the effective length of the rings 
considered here is very small compared to the one of the lattice rings 
studied in \cite{orlandinistella}, and no simple argument is presently available to predict if
the $\alpha \approx 0.7$ exponent is expected to be maintained
for increasing contour lengths where the size of the knotted portion 
is supposed to be negligible with respect to the one of the whole ring.

In conclusion, the above results provide a vivid picture of the 
key features of the dynamics of the knotted region.
Specifically, they illustrate that the displacement of the knot center of mass follows
``passively'' the one of the whole ring because of the very long times
required by the knot to slide along the ring.

\section{Summary and Conclusions}\label{sec:concl}

We reported on a systematic computational study of the equilibrium and
dynamics of solutions of unconcatenated ring polymers with different knot topology.
Specifically, molecular dynamics simulations
(with no explicit treatment of the solvent)
at fixed-volume and constant temperature were carried out on bead-spring
models of semiflexible unknotted and trefoil-knotted rings for several solution densities.
This framework was used to explore the extent to
which the interplay between topological constraints (knotting) and the
geometrical self (intra-chain) and mutual (inter-chain) entanglements
affect both the equilibrium and the dynamical properties of the rings.
The study complements previous investigations of dense solutions of
unlinked and unknotted ring polymers of various contour lengths.

Regarding the equilibrium metric properties it is found that changes
of the inter-chain or intra-chain entanglement operated by varying the
solution density and/or ring topology, affect modestly the average
ring size and shape compared to the infinitely-diluted case.
Specifically, the root mean radius of gyration of both unknotted and
trefoil rings at the highest solution densities (occupied volume
fraction equal to 0.4) are at most about $40 \%$ smaller than in the
unconstrained case. At all densities, trefoils are smaller (between
10\% and 20\% in linear size) and slightly more globular than unknots.
Yet the average exposed surface of rings with different topology is
practically the same and about constant at all densities.  These
results offer an interesting insight regarding the compactness of the
rings and, in particular, they indicate that the moderate decrease in
ring size following the increasing intra- and inter-chain entanglement
does not preclude the persistence of voids and cavities within the
rings convex hull so that the exposed surface area is about the same
as for unconstrained isolated rings.

The weak dependence of the overall ring geometric features on solution
density prompts the question of whether, for knotted rings, the metric
properties of the knotted ring portion are weakly affected too. By
knotted ring portion we refer to the shortest arc that accommodates
the knot. This question is addressed by using the effective and
transparent knot-location algorithm that was used recently to show
that in isolated rings the length of the knotted region varies
dramatically with increasing spherical confinement. At variance with
the latter case, it is found that the length of the average size of
the knotted ring portion is practically insensitive to density
variations.

While ring equilibrium metric properties are only weakly affected
by variations of solution density and ring topology, the opposite
holds for kinetic properties. The characteristic times of ring size
relaxation, reorientation, and center of mass diffusion change by one order of magnitude across the considered density range.

Further topology-dependent aspects of ring kinetics were highlighted
by monitoring various kinetic observables of the ring knotted portion.
Specifically we focused on how such region displaces in space and
along the ring contour. The backbone motion is found to follow
standard one-dimensional diffusion and the corresponding diffusion
coefficient is smaller (much smaller at low concentrations) than the
one of the ring center of mass. Consequently, for time scales over
which a ring moves appreciably in space, its knotted portion remains
practically quenched and, as we verified, diffuses in space as any
other equally-long tagged portion of the ring.

In summary, the results presented here offer novel insights into the
impact of intra-chain and inter-chain entanglement in solutions of
ring polymers. The mild dependence of equilibrium metric properties on
both the above-mentioned effects has no parallel with the behavior of
dense systems of rings obtained by three-dimensional spatial (spherical) confinement.
This suggests that an interesting novel avenue
to address in future work would be to examine analogous effects for
dense systems of rings obtained by two-and one-dimensional confinement
(slabs and channels), which are increasingly adopted for advanced
polymer micromanipulations. By converse, the sensitive dependence of
various ring kinetic properties on solution density and ring topology
suggests that both effects should be relevant to the studies of rheology
and transport properties in semi-dilute solutions of circular polymers.

\section{Acknowledgments}
Numerical calculations were partly carried out in CINECA,
and partly under the HPC-Europa2 project (project number: 228398)
with the support of the European Commission Capacities Area-Research Infrastructures Initiative.
This work made use of the facilities of HECToR, the UK's national high
performance computing service, which is provided by UoE HPCx Ltd at
the University of Edinburgh, Cray Inc and NAG Ltd, and funded by the
Office of Science and Technology through EPSRC's High End Computing Programme.

\vspace{10mm}

\noindent Reprinted with permission from\\
A. Rosa et al., {\it Macromolecules}, 2011, {\bf 44} (21), pp. 8668-8680.\\
Copyright (2011) American Chemical Society.

\newpage 

%

\setcounter{figure}{0}

\renewcommand{\figurename}{SUPPORTING FIG.}
\newpage 
\begin{figure*}
\includegraphics[width=5.0in]{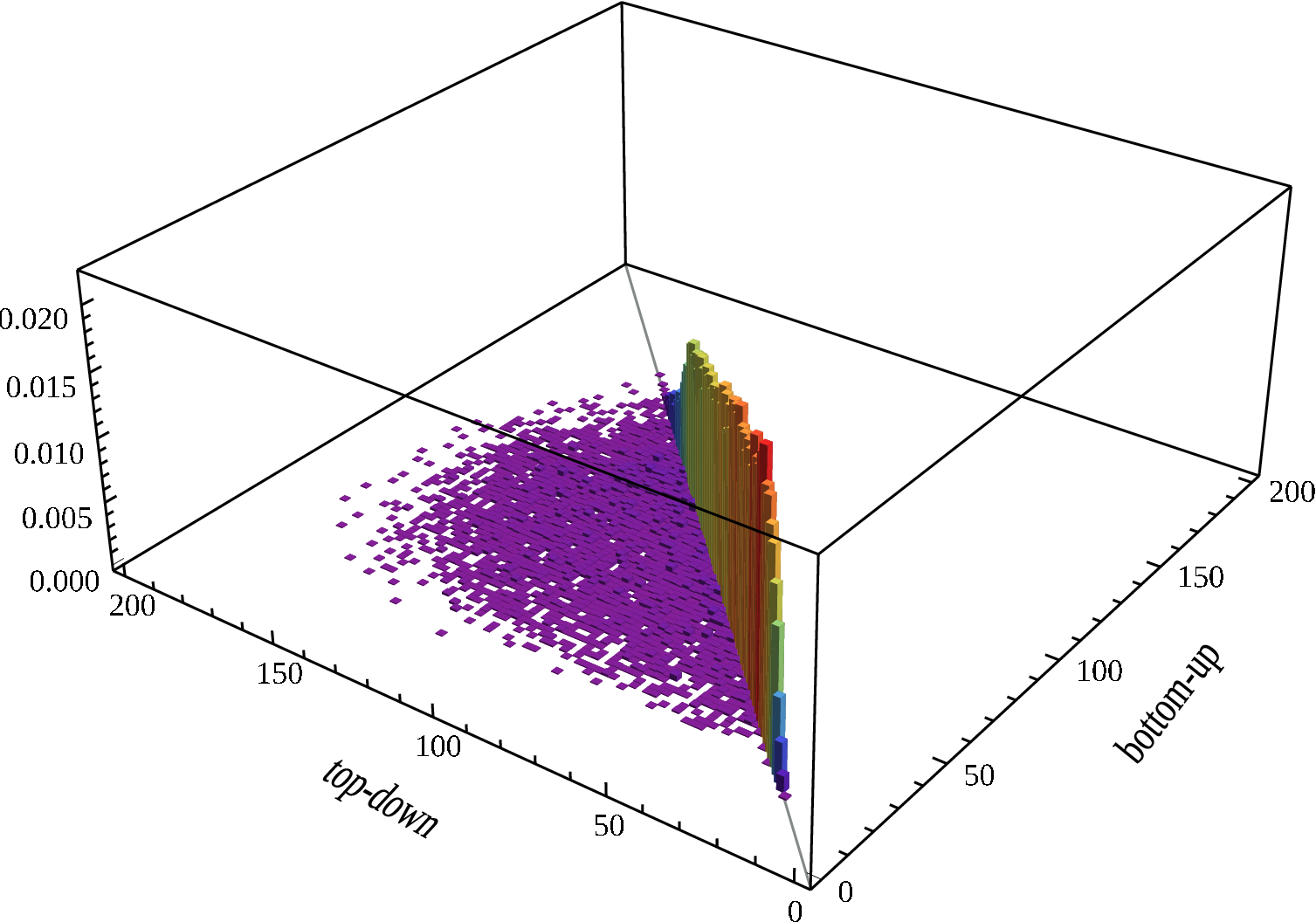}
\caption{
\label{supplfig:topdownVSbottomup}
Three-dimensional histogram comparing top-down and bottom-up search schemes
of the length, $L_{knot}$ of the knotted portion of the chain (see Methods, Sec. IID):
the histogram reports the number of times the pair of values given by $\left( L_{knot}^{bottom-up}, L_{knot}^{top-down} \right)$
occur in the course of the simulation.
It is evident that the two algorithms measure the same knotted portion in the majority of cases.
The case shown corresponds to ring polymers at monomer concentration $\rho = 0.4$.
Analogous results are found at the other concentrations.
}
\end{figure*}

\end{document}